\documentclass[preprint,amsmath,amssymb,aps,prd,a4paper]{revtex4-2}

\usepackage{graphicx}
\usepackage{bm}
\usepackage{cleveref}

\usepackage{xcolor}
\usepackage[normalem]{ulem}

\begin{document}

\title{Leading-twist to higher-twist generalized parton distributions of the pseudoscalar mesons at non-zero skewness}

\author{Abi Jebarson A}
\email{abijebarson@gmail.com}

\author{Navpreet Kaur}
\email{knavpreet.hep@gmail.com}

\author{Harleen Dahiya}
\email{dahiyah@nitj.ac.in}

\affiliation{Department of Physics, Dr.~B.~R.~Ambedkar National Institute of Technology, Jalandhar, Punjab, 144008, India }

\begin{abstract}
    We investigate the multidimensional partonic structure of spin-0 mesons, specifically the pion and the kaon, by evaluating their complete set of eight generalized parton distributions (GPDs) up to twist-4. Utilizing the light-front quark model (LFQM) with the Brodsky-Huang-Lepage (BHL) prescription, we compute these distributions in the kinematically rich non-zero skewness ($\xi \neq 0$) domain, strictly within the DGLAP region, $x \in [\xi, 1]$. To construct a three-dimensional tomographic picture, we perform Fourier transforms of the momentum-space GPDs to obtain the impact parameter dependent parton distribution functions (IPDPDFs) in the transverse plane and the corresponding diffraction patterns in the longitudinal coordinate space. The numerical results explicitly reveal the consequences of $\mathrm{SU}(3)$ flavor symmetry breaking in the LFQM, as the strange quark in the kaon dynamically shifts spatial localizations compared to the lighter up quarks. We also observe that while higher-twist correlations exhibit massive amplitude scaling in the pion, they are heavily suppressed by the larger macroscopic mass of the kaon.
\end{abstract}

\maketitle

\section{Introduction}

One of the fundamental goals of quantum chromodynamics (QCD) is to understand the complex internal structure of hadrons in terms of their constituent valence quarks, gluons, and sea quarks. While QCD is highly successful in predicting experimentally verifiable phenomena at high energy scales where the strong coupling is weak and perturbative calculations are possible, describing the bound-state structure of a hadron requires navigating the non-perturbative regime of the theory. Traditionally, these non-perturbative dynamics have been encoded in one-dimensional parton distribution functions (PDFs)~\cite{Feynman:1969ej, Bjorken:1969ja, Collins:1981uw,Martin:2012da, Jaffe:1991ra, Martin:2009iq} and elastic form factors~\cite{Hofstadter:1956qs, Perdrisat:2006hj, Diehl:2003ny, Diehl:2004cx, Khodjamirian:2006st}. However, to completely understand the internal structure of hadrons, one must move beyond the conventional one-dimensional representations. Generalized parton distributions (GPDs)~\cite{Muller:1994ses, Radyushkin:1997ki, Ji:1996nm, Diehl:2003ny, Diehl:2015uka, Boffi:2007yc, Radyushkin:1996nd, Belitsky:2001ns, Belitsky:2005qn, Diehl:2013xca} have emerged as universal and powerful tools to achieve this, providing a comprehensive, three-dimensional framework that embodies both the spatial and momentum distributions of partons.

Among hadrons, the pion and kaon, discovered around eighty years ago~\cite{Rochester:1947mi, Lattes:1947mw}, play a unique and pivotal role in QCD. The pion is the lightest bound state in nature and is identified as the Nambu-Goldstone boson associated with the spontaneous breaking of chiral symmetry~\cite{Nambu:1961tp}. Similarly, the kaon is also a Nambu-Goldstone boson. However, the presence of the heavier strange quark explicitly breaks the $\mathrm{SU}(3)$ flavor symmetry~\cite{Pich:1995bw}. Studying the kaon structure alongside the pion is highly compelling because the unequal masses of the constituent quarks introduce an inherent asymmetry into the system~\cite{SKaur:2019jow, Kaur:2020vkq, Kaur:2020par}. Comparing the multidimensional structural observables of these mesons provides a direct window into the fundamental symmetry-breaking mechanisms of the strong interaction.

The theoretical description of these spin-0 mesons is remarkably rich. The fully unintegrated off-diagonal quark-quark correlator for a spin-0 hadron can be parameterized to yield a complete classification of GPDs up to higher twists~\cite{Meissner:2008ay, Signal:1996ct, Avakian:2010br, Lorce:2014hxa, Lu:2012gu, Mao:2013waa, Mao:2014aoa, Liu:2021ype}. In total, there are eight GPDs for a spin-0 hadron. While extensive research has focused on the zero-skewness limit ($\xi = 0$), where the initial and final longitudinal momenta of the hadron are equal, exploring the non-zero skewness region ($\xi \neq 0$) is crucial for a complete understanding of exclusive processes. This pursuit is driven both by foundational theoretical frameworks~\cite{Diehl:2003ny, Ji:1996nm, Kumericki:2016ehc, Kroll:2014tma, Anikin:2000em} and by robust experimental mapping efforts at facilities including Jefferson Lab, HERMES, and ZEUS~\cite{JeffersonLabHallA:2007jdm, Defurne:2017paw, Dupre:2016mai, CLAS:2004cri, ZEUS:2005bhf, HERMES:2007hrc}. GPDs at non-zero skewness encode the longitudinal momentum transfer between the initial and final states, providing significantly more information about the internal partonic dynamics than their zero-skewness counterparts. Specifically, characterizing this explicit $\xi$-dependence captures the off-diagonal momentum correlations between initial and final state partons, which a zero-skewness GPD cannot capture. This procedure uniquely unveils the diffraction-like longitudinal spatial localization of partons along the direction of hadron motion. Furthermore, the specific GPDs that are constrained by hermiticity to strictly vanish at zero skewness (such as $F_2$ and $H_2$) can only be activated and accessed through non-zero skewness calculations, making this kinematically rich domain essential for a truly comprehensive mapping of hadron structure.

Beyond their momentum-space definitions, GPDs act as theoretical lenses to view the spatial distribution of partons through Fourier transforms. Taking the Fourier transforms of GPDs with respect to the transverse momentum transfer $\Delta_\perp$ yields the impact parameter dependent parton distribution functions (IPDPDFs)~\cite{Diehl:2002he, Burkardt:2002hr, Burkardt:2000za, Burkardt:2000uu}. This provides a probabilistic interpretation of finding a parton with a longitudinal momentum fraction $x$ at a transverse distance $b_\perp$ from the transverse center of momentum. Furthermore, a Fourier transform with respect to the skewness parameter $\xi$ projects the GPDs into a longitudinal boost-invariant space $\sigma$, revealing the diffraction patterns of the target hadron~\cite{Brodsky:2006ku, Collins:2011ca, Meissner:2007rx}.

Because calculating GPDs directly from first principles in QCD remains formidably difficult, theoretical investigations have long relied on a diverse range of effective phenomenological models~\cite{Ji:1997gm, Scopetta:2002xq, Scopetta:2003et, Boffi:2002yy, Boffi:2003yj, Pasquini:2006dv, Mondal:2015uha, Gutsche:2014zua, Kaur:2018ewq, Kaur:2018ewq, Theussl:2002xp, Tiburzi:2002sw} to capture the non-perturbative structure of light mesons. However, this landscape is rapidly evolving. While traditionally hindered by the impossibility of directly simulating non-local light-cone correlators on a Euclidean spacetime lattice, the advent of Large-Momentum Effective Theory (LaMET)~\cite{Ji:2013dva} and related pseudo-distribution approaches has opened the door to first-principles extractions of $x$-dependent parton physics. Recent breakthroughs in lattice QCD have successfully applied these frameworks to compute the valence quasi-GPDs and three-dimensional impact-parameter spatial distributions of the pion~\cite{Chen:2019lcm, Ding:2024saz}. Complementing these first-principles advancements, phenomenological efforts continue to expand. For instance, recent works have evaluated leading twist-2 GPDs within nonlocal chiral quark frameworks~\cite{Son:2024uet} and another work explored off-shell variations of pion distributions~\cite{Zhang:2025xtn}. Concurrently, investigations into the higher-twist domain have advanced significantly. These include detailed light-front and spectator model formalisms~\cite{Lorce:2016ugb, Pasquini:2018oyz, Mukherjee:2002pq, Mukherjee:2002xi, Balla:1997hf, Dressler:1999hc}, chiral and soliton model approaches~\cite{Schweitzer:2003uy, Wakamatsu:2003uu, Wakamatsu:2007nc, Ohnishi:2003mf, Cebulla:2007ej}, and more recent phenomenological developments in both mesons and baryons~\cite{Sharma:2023wha, Zhu:2023lst, Aslan:2018tff, Sharma:2023tre, Bhattacharya:2023nmv, Zhang:2023xfe, Sharma:2022ylk, Aslan:2019jis}. Building upon these efforts, Luan and Lu recently utilized a light-cone quark model to evaluate the twist-3 and twist-4 valence GPDs of the pion and kaon~\cite{Luan:2024dvc}, where the calculations were limited to the zero-skewness limit ($\xi = 0$).

In this vanishing skewness limit, the square of the invariant momentum transfer, $-t$, simplifies uniquely to the square of the transverse momentum transfer, $-t = \Delta_\perp^2$, facilitating a straightforward tomographic mapping to the impact parameter space. Conversely, in the non-zero skewness regime ($\xi \neq 0$), $-t$ acquires a longitudinal component coupled to both the skewness and the macroscopic hadron mass $M$, governed by the kinematic relation $-t = (\Delta_\perp^2 + 4\xi^2 M^2)/(1-\xi^2)$. While prior modeling efforts have provided valuable insights into hadronic structure and form factors~\cite{Puhan:2025pfs, Brodsky:2000ii, Lepage:1980fj, Bacchetta:2008af, Lu:2006kt, Luan:2022fjc, Ma:2018ysi, Luan:2024nwc, Luan:2023lmt}, a simultaneous numerical mapping of the complete set of spin-0 GPDs~\cite{Meissner:2008ay} across the kinematically richer non-zero skewness domain has remained largely unexplored.

In this paper, we bridge this gap by calculating the complete set of eight GPDs, up to twist-4, for the valence quarks of the pion and the kaon at non-zero skewness. The leading-twist GPDs have been calculated previously~\cite{Kaur:2018ewq, SKaur:2019jow} and are included here for completeness. To evaluate these distributions, we employ the light-front quark model (LFQM) utilizing the Brodsky-Huang-Lepage (BHL) prescription~\cite{Brodsky:1982nx, Xiao:2003wf, Qian:2008px} for constructing the light-front wave functions (LFWFs). The BHL prescription provides a robust phenomenological framework, effectively capturing the transverse momentum dependence through a Gaussian wave function while satisfying fundamental theoretical constraints. While this phenomenological approach captures essential kinematic and orbital angular momentum correlations, it does not incorporate the dynamical effects of QCD mass generation~\cite{Zhang:2020gaj, Roberts:2021nhw, Raya:2021zrz}. Utilizing the overlap representation within this light-front framework, we evaluate the correlators. It is important to note that restricting the Fock state expansion to the valence $q\bar{q}$ sector implicitly applies a Wandzura-Wilczek-like approximation~\cite{Wandzura:1977qf} for higher-twist distributions, as explicit quark-gluon-quark ($q\bar{q}g$) correlations stemming from the transverse gauge link are necessarily truncated. We present comprehensive three-dimensional surface plots depicting the behavior of these GPDs as functions of $x$ and $-t$ for fixed $\xi$, as well as functions of $x$ and $\xi$ for fixed $-t$. Furthermore, to extract an intuitive spatial picture, we numerically Fourier transform the GPDs to evaluate the IPDPDFs in the transverse plane (with $b_\perp$ expressed in femtometers) and the corresponding diffraction patterns in the longitudinal space $\sigma$.

This paper is organized as follows. In Sec.~II, we introduce the theoretical formalism of the light-front framework and detail the specific LFQM employed in this study. In Sec.~III, we systematically define the complete set of eight meson GPDs, outlining the relevant correlators from twist-2 up to twist-4. The mathematical formalism for projecting these momentum-space distributions into coordinate space via Fourier transforms, yielding both the IPDPDFs and the longitudinal position space distributions, is established in Sec.~IV. In Sec.~V, we present and extensively discuss our numerical results for the GPDs and their corresponding spatial tomographies. Finally, a summary of our findings is provided in Sec.~VI.

\section{Model description}
\subsection{Light-front framework}

In a light-front frame~\cite{Dirac:1949cp, Brodsky:1997de}, a generic four-vector is defined by its components as $a = [a^+, a^-, \bm{a}_\perp]$, where the longitudinal light-cone components are given by $a^\pm = a^0 \pm a^3$, and $\bm{a}_\perp = (a^1, a^2)$ is the transverse component.

A hadronic system $\mathcal{H}$ can be expressed in terms of its constituents as a Fock state. The light-front Fock state expansion of a hadronic eigenstate with total momentum $P$ and spin projection $S_z$ is given by~\cite{Qian:2008px, Brodsky:2000xy}
\begin{align}
    |\mathcal{H} (P^+, \bm{P}_\perp, S_z) \rangle & = \sum_{n,\lambda_i} \int \prod_{i=1}^n \frac{\mathrm{d} x_i \mathrm{d}^2 \bm{k}_{\perp i}}{\sqrt{x_i}~16\pi^3} \nonumber         \\
                                                  & \times 16\pi^3 \delta\Big(1-\sum_{i=1}^n x_i\Big)\delta^{(2)}\Big(\sum_{i=1}^n \bm{k}_{\perp i}\Big) \nonumber                    \\
                                                  & \times | n; x_i P^+, x_i \bm{P}_\perp+\bm{k}_{\perp i}, \lambda_i \rangle \psi_{n/\mathcal{H}}^{\lambda_i}(x_i,\bm{k}_{\perp i}),
    \label{eq:hadron}
\end{align}
where $x_i = k_i^+ / P^+$ is the longitudinal momentum fraction of the $i^{\text{th}}$ constituent parton, $\bm{k}_{\perp i}$ is its intrinsic transverse momentum, $\lambda_i$ is its helicity, and $\psi_{n/\mathcal{H}}^{\lambda_i}$ is the LFWF amplitude.

The light-front four-momenta of the initial ($P'$) and final ($P''$) hadron states are evaluated in a symmetric Drell-Yan-West frame as
\begin{align}
    P'  & = \bigg[(1+\xi)P^+, \frac{M^2+\bm{\Delta}^2_\perp/4}{(1+\xi)P^+},\frac{\bm{\Delta}_\perp}{2} \bigg],  \\
    P'' & = \bigg[(1-\xi)P^+, \frac{M^2+\bm{\Delta}^2_\perp/4}{(1-\xi)P^+},-\frac{\bm{\Delta}_\perp}{2} \bigg],
\end{align}
where $M$ is the macroscopic mass of the hadron, $\xi$ is the skewness parameter representing the longitudinal momentum transfer fraction, and $\bm{\Delta}_\perp$ is the transverse momentum transfer. The average four-momentum of the meson $P^\mu = \frac{1}{2}(P' + P'')^\mu$ and the total four-momentum transfer $\Delta^\mu = P'^\mu - P''^\mu$ are given by
\begin{align}
    P      & = \bigg[P^+, \frac{M^2+\bm{\Delta}_\perp^2/4}{(1-\xi^2)P^+}, \bm{0}_\perp \bigg],                       \\
    \Delta & = \bigg[2 \xi P^+, -\frac{\xi \bm{\Delta}_\perp^2 + 4 \xi M^2}{2(1-\xi^2)P^+}, \bm{\Delta}_\perp\bigg].
\end{align}
The momentum of the $i^{\text{th}}$ constituent parton is defined as
\begin{align}
    k_i & = \bigg[x_i P^+, \frac{(\bm{k}_{\perp i} + x_i \bm{P}_\perp)^2+m^2_i}{x_i P^+}, \bm{k}_{\perp i} + x_i \bm{P}_\perp\bigg],
\end{align}
where $m_i$ is the mass of the $i^{\text{th}}$ constituent parton. The $n$-particle Fock states $|k_i^+, \bm{k}_{\perp i}, \lambda_i \rangle$ are normalized as follows,
\begin{align}
    \langle n; k'^+_i, \bm{k}'_{\perp i}, \lambda'_i | n; k^+_i, \bm{k}_{\perp i}, \lambda_i \rangle & = \prod_{i=1}^n 16 \pi^3 k_i^+ \delta(k'^+_i - k_i^+) \nonumber                          \\
                                                                                                     & \times \delta^{(2)}(\bm{k}'_{\perp i} - \bm{k}_{\perp i}) \delta_{\lambda'_i \lambda_i}.
\end{align}

For a meson ($n=2$), taking the active quark momentum fraction as $x$ and the intrinsic transverse momentum as $\bm{k}_\perp$, and the helicities of the active quark and spectator antiquark (or active antiquark and spectator quark) as $\lambda_1$ and $\lambda_2$ respectively, the two-particle Fock state expansion in Eq.~\eqref{eq:hadron} reduces to~\cite{Xiao:2003wf, Qian:2008px}
\begin{align}
    |\mathcal{M}(P,S)\rangle & = \sum_{\lambda_1, \lambda_2} \int \frac{dx d^2\bm{k}_\perp}{\sqrt{x(1-x)} 16\pi^3} |x, \bm{k}_\perp, \lambda_1, \lambda_2 \rangle \psi^{\lambda_1,\lambda_2}_{S_z}(x, \bm{k}_\perp).
    \label{eq:meson}
\end{align}

\subsection{Light-front quark model}
The total two-body LFWF for a generic pseudoscalar meson $\mathcal{M}$ (where $\mathcal{M} = \pi, K$) is expressed as a product of a momentum space wave function $\varphi^\mathcal{M}$ and a spin-flavor wave function $\chi^\mathcal{M}$~\cite{Qian:2008px}
\begin{align}
    \psi_{S_{z}}^{\mathcal{M}}\left(x, \bm{k}_\perp, \lambda_{1}, \lambda_{2}\right) = \varphi^\mathcal{M}\left(x, \bm{k}_\perp\right) \chi_{S_{z}}^{\mathcal{M}}\left(x, \bm{k}_\perp, \lambda_{1}, \lambda_{2}\right).
\end{align}
Let $m_1$ represent the mass of the active constituent quark and $m_2$ represent the mass of the spectator antiquark. The spin-flavor wave function $\chi^\mathcal{M}$ is constructed by transforming the ordinary equal-time spin states into light-front helicity states via the Melosh-Wigner rotation~\cite{Xiao:2003wf, Qian:2008px}. To streamline the notation for the elements of this rotation matrix, we define the constituent mass combination $m_{12} = (1-x)m_1 + x m_2$. For a generic pseudoscalar meson $\mathcal{M}$ with spin projection $S_z = 0$, the explicit helicity configurations generated with all possible combinations of $\uparrow$ and $\downarrow$ for $\lambda_1$ and $\lambda_2$, by the Melosh rotation are~\cite{Xiao:2003wf, SKaur:2019jow}
\begin{align}
    \psi^{\mathcal{M}}_{0}(x, \bm{k}_\perp, \uparrow, \downarrow)   & = +\frac{m_{12}}{\sqrt{2(\bm{k}_\perp^2 + m_{12}^2)}} \varphi^{\mathcal{M}} \quad (l^z = 0), \nonumber       \\
    \psi^{\mathcal{M}}_{0}(x, \bm{k}_\perp, \downarrow, \uparrow)   & = -\frac{m_{12}}{\sqrt{2(\bm{k}_\perp^2 + m_{12}^2)}} \varphi^{\mathcal{M}} \quad (l^z = 0), \nonumber       \\
    \psi^{\mathcal{M}}_{0}(x, \bm{k}_\perp, \uparrow, \uparrow)     & = -\frac{k_x - i k_y}{\sqrt{2(\bm{k}_\perp^2 + m_{12}^2)}} \varphi^{\mathcal{M}} \quad (l^z = -1), \nonumber \\
    \psi^{\mathcal{M}}_{0}(x, \bm{k}_\perp, \downarrow, \downarrow) & = -\frac{k_x + i k_y}{\sqrt{2(\bm{k}_\perp^2 + m_{12}^2)}} \varphi^{\mathcal{M}} \quad (l^z = +1),
\end{align}
where $\bm{k}_\perp = (k_x, k_y)$.

Following the BHL prescription, the momentum space wave function $\varphi^\mathcal{M}$ is modeled as a phenomenological Gaussian distribution dependent upon the invariant mass squared of the free constituent system, $M_0^2 = \frac{\bm{k}_\perp^2 + m_1^2}{x} + \frac{\bm{k}_\perp^2 + m_2^2}{1-x}$ \cite{Xiao:2003wf, Qian:2008px},
\begin{align}
    \varphi^{\mathcal{M}}(x, \bm{k}_\perp) & = A^{\mathcal{M}} \exp \left[ -\frac{1}{8 \beta_{\mathcal{M}}^2} \left( \frac{\bm{k}_\perp^2 + m_1^2}{x} + \frac{\bm{k}_\perp^2 + m_2^2}{1-x} \right) - \frac{(m_1^2 - m_2^2)^2}{8 \beta_{\mathcal{M}}^2 \left(\frac{\bm{k}_\perp^2 + m_1^2}{x} + \frac{\bm{k}_\perp^2 + m_2^2}{1-x}\right)} \right],
\end{align}
where $A^{\mathcal{M}}$ is the normalization constant and $\beta_\mathcal{M}$ is a harmonic oscillator parameter that dictates the transverse momentum spread. Notice that for the pion ($\mathcal{M} = \pi$), flavor symmetry implies equal constituent masses ($m_1 = m_2 = m$). Under this condition, the second term in the exponential naturally vanishes, recovering the standard symmetric Gaussian form. For the kaon ($\mathcal{M} = K$), explicit $\mathrm{SU}(3)$ flavor symmetry breaking dictates $m_1 \neq m_2$, activating the full functional structure to maintain physical endpoint behaviors.

\section{Generalized parton distributions}

The GPDs are rigorously defined via the unintegrated off-forward quark-quark correlator~\cite{Meissner:2008ay, Diehl:2003ny}
\begin{align}
    F^{[\Gamma]}(x, \xi, t) = \frac{1}{2} \int \frac{d z^{-}}{2 \pi} e^{i k.z} \left. \langle P'' | \bar{\psi}\Big(-\frac{z}{2}\Big) \Gamma \mathcal{W}\Big(-\frac{z}{2}, \frac{z}{2}\Big) \psi\Big(\frac{z}{2}\Big) | P' \rangle \right|_{z^+ = 0, \bm{z}_\perp = \bm{0}_\perp},
\end{align}
where $z$ is the light-like separation distance, $\mathcal{W}$ is the Wilson line ensuring gauge invariance and is assumed unity for this calculation, and $\Gamma$ spans the Dirac matrices governing different twist structures. In this formalism, $x$ represents the average light-cone momentum fraction of the active quark, while $\xi$ is the skewness parameter defining the longitudinal momentum transfer.

The evaluation of GPDs via the overlap of two-body light-front Fock states is strictly valid only in the DGLAP (Dokshitzer-Gribov-Lipatov-Altarelli-Parisi) kinematic region, defined by $x > \xi \ge 0$~\cite{Diehl:2000xz, Brodsky:2000xy, Sharma:2023ibp}. In this domain, the process corresponds to the emission of a quark with momentum fraction $x+\xi$ and its reabsorption with a positive fraction $x-\xi$. The methodology does not inherently cover the ERBL (Efremov-Radyushkin-Brodsky-Lepage) region ($-\xi < x < \xi$), which would require non-diagonal particle number transitions. It is important to emphasize that restricting the light-front Fock state expansion strictly to the valence $q\bar{q}$ sector is equivalent to applying a Wandzura-Wilczek-type approximation for higher-twist distributions \cite{Wandzura:1977qf}. In this approximation, the dynamical higher-twist effects that come from explicit quark-gluon-quark ($q\bar{q}g$) correlations via the transverse gauge link are neglected. Consequently, the twist-3 and twist-4 distributions presented here capture only the kinematic higher-twist contributions arising from the transverse momentum and orbital motion of the valence quarks. In this work, we focus exclusively on the DGLAP region and provide results corresponding to the hadronic model scale \cite{Wang:2024wny, Yin:2023dbw}, which is fixed at $Q_0^2 = 0.20 \text{ GeV}^2$ based on our previous evaluations within this specific framework~\cite{A:2025sgk}. These results can subsequently serve as initial boundary conditions for DGLAP evolution to higher experimental $Q^2$ scales.

To formalize the kinematics within this overlap representation, the active quark longitudinal momentum fractions relative to the initial and final meson states are defined as $x' = (x+\xi)/(1+\xi)$ and $x'' = (x-\xi)/(1-\xi)$, respectively. This mapping critically ensures that the individual momentum fractions satisfy $x', x'' \in [0,1]$, which is the defining physical characteristic of the DGLAP kinematic domain.

The individual GPDs are parameterized from the correlators for a pseudoscalar meson as follows~\cite{Meissner:2008ay},
\begin{align}
    F^{[\gamma^+]}               & = F_1^\mathcal{M}(x, \xi, t), \label{eq:gpd_F1}                                                                                  \\
    F^{[i\sigma^{j+}\gamma_5]}   & = \frac{i\epsilon_\perp^{ij}\Delta_\perp^i}{M} H_1^\mathcal{M}(x, \xi, t), \label{eq:gpd_H1}                                     \\
    F^{[\mathbf{1}]}             & = \frac{M}{P^+} E_2^\mathcal{M}(x, \xi, t), \label{eq:gpd_E2}                                                                    \\
    F^{[\gamma^j]}               & = \frac{M}{P^+} \left[ \frac{\Delta_\perp^j}{M} F_2^\mathcal{M}(x, \xi, t) \right], \label{eq:gpd_F2}                            \\
    F^{[\gamma^j \gamma_5]}      & = \frac{M}{P^+} \left[ \frac{i \epsilon_\perp^{ij} \Delta_\perp^i}{M} G_2^\mathcal{M}(x, \xi, t) \right], \label{eq:gpd_H2}      \\
    F^{[i \sigma^{ij} \gamma_5]} & = \frac{M}{P^+} \left[ i \epsilon_\perp^{ij} H_2^\mathcal{M}(x, \xi, t) \right], \label{eq:gpd_G2}                               \\
    F^{[\gamma^-]}               & = \frac{M^2}{(P^+)^2} F_3^\mathcal{M}(x, \xi, t), \label{eq:gpd_F3}                                                              \\
    F^{[i \sigma^{j-} \gamma_5]} & = \frac{M^2}{(P^+)^2} \left[ \frac{i \epsilon_\perp^{ij} \Delta_\perp^i}{M} H_3^\mathcal{M}(x, \xi, t) \right],\label{eq:gpd_H3}
\end{align}
where $\sigma^{i j}=i\left[\gamma^{i}, \gamma^{j}\right] / 2 $ and $\epsilon_{\perp}^{i j}=\epsilon^{-+i j}$. Note that $M$ here explicitly denotes the macroscopic meson mass, serving to balance the mass dimension of the form factors.

The active quark intrinsic transverse momenta in the initial ($\bm{k}'_\perp$) and final ($\bm{k}''_\perp$) states are kinematically linked to the average intrinsic transverse momentum $\bm{k}_\perp$ and the transverse momentum transfer $\bm{\Delta}_\perp$. Because the spectator quark momentum remains unchanged during the interaction, the intrinsic transverse momenta safely map as
\begin{align}
    \bm{k}'_\perp  & = \bm{k}_\perp + (1-x')\frac{\bm{\Delta}_\perp}{2}, \nonumber \\
    \bm{k}''_\perp & = \bm{k}_\perp - (1-x'')\frac{\bm{\Delta}_\perp}{2}.
\end{align}
The invariant momentum transfer squared is defined as $t = \Delta^2$, which yields the transverse magnitude $\Delta_\perp = |\bm{\Delta}_\perp| = \sqrt{-(1-\xi^2)t - 4M^2\xi^2}$. For a physical scattering process, the requirement that $\bm{\Delta}_\perp^2 \ge 0$ imposes a strict kinematic lower bound on the magnitude of the space-like momentum transfer,
\begin{equation}
    -t \ge (-t)_{\text{min}} = \frac{4M^2\xi^2}{1-\xi^2}.
\end{equation}
We define the transverse momentum transfer vector explicitly as $\bm{\Delta}_\perp = (\Delta_x, \Delta_y)$.

We define the helicity overlap matrix elements $\mathcal{A}_{\lambda'\lambda}$, where $\lambda$ and $\lambda'$ denote the initial and final helicity states of the active quark, by tracing over the unobserved spectator quark helicity $\lambda_s$,
\begin{equation}
    \mathcal{A}_{\lambda'\lambda} = \sum_{\lambda_s \in \{\uparrow, \downarrow\}} \big( \psi_{f}^{\lambda' \lambda_s} \big)^* \psi_{i}^{\lambda \lambda_s}.
\end{equation}
To compress the analytical forms of the correlators, we project these overlap amplitudes into a scalar $S$, a pseudoscalar $P$, and a 2D vector $\bm{V}$,
\begin{align}
    S      & = \mathcal{A}_{\uparrow\uparrow} + \mathcal{A}_{\downarrow\downarrow},                                                                                       \\
    P      & = \mathcal{A}_{\uparrow\uparrow} - \mathcal{A}_{\downarrow\downarrow},                                                                                       \\
    \bm{V} & = \big( \mathcal{A}_{\uparrow\downarrow} + \mathcal{A}_{\downarrow\uparrow}, \ i(\mathcal{A}_{\downarrow\uparrow} - \mathcal{A}_{\uparrow\downarrow}) \big).
\end{align}
To capture the complex phase space routing in the higher-twist correlators, we define two auxiliary transverse momentum vectors $\bm{A}$ and $\bm{B}$, representing interference between the intrinsic transverse momentum $\bm{k}_\perp$ and the momentum transfer $\bm{\Delta}_\perp$,
\begin{align}
    \bm{A} & = 2\xi \bm{k}_\perp - x\bm{\Delta}_\perp, \\
    \bm{B} & = 2x \bm{k}_\perp - \xi\bm{\Delta}_\perp.
\end{align}
For generic 2D vectors $\bm{a}$ and $\bm{b}$, the cross product yields a scalar defined as $\bm{a} \times \bm{b} = a_x b_y - a_y b_x$. We also define the orthogonal $z$-axis rotation and a rank-2 symmetric tensor contraction $\mathcal{T}$ for a generic vector $\bm{a}$,
\begin{align}
    \hat{\bm{z}} \times \bm{a}  & = (-a_y, a_x),                                 \\
    \mathcal{T}(\bm{a}, \bm{V}) & = 2(\bm{a}\cdot\bm{V})\bm{a} - \bm{a}^2\bm{V}.
\end{align}
Note that in the subsequent sections, boldface vector correlators such as $\bm{F}^{[\gamma^\perp]}$ or $\bm{F}^{[i\sigma^{\perp+}\gamma_5]}$ are used strictly as a compressed 2D shorthand to simultaneously represent the grouped transverse components (e.g., $j=1, 2$). They should not be confused with the individual Lorentz component parameterizations $F^{[\gamma^j]}$ introduced earlier.

\subsection{Twist-2 Correlators}
The leading-twist correlators scale with $\mathcal{N}_2$ and extract the pure scalar and transverse vector components directly. We define the kinematic coefficient,
\begin{align}
    \mathcal{N}_2 = \frac{\sqrt{x^2-\xi^2}}{16 \pi^3 x}.
\end{align}
This gives the two twist-2 GPDs $F_1^\mathcal{M}$ and $H_1^\mathcal{M}$ from their correlators,
\begin{align}
    F^{[\gamma^+]}                      & = \int d^2\bm{k}_\perp \, \mathcal{N}_2 S,\label{eq:gpd_F1_dec} \\
    \bm{F}^{[i\sigma^{\perp+}\gamma_5]} & = \int d^2\bm{k}_\perp \, \mathcal{N}_2 \bm{V}.
    \label{eq:gpd_H1_dec}
\end{align}

\subsection{Twist-3 Correlators}
The sub-leading twist-3 correlators introduce dynamical mixing scaled by $\mathcal{N}_3$. Mass terms couple to the primary projections, while momentum components induce orbital angular momentum interference. We define the kinematic coefficient,
\begin{align}
    \mathcal{N}_3 = \frac{1}{32 \pi^3 x P^+ \sqrt{x^2-\xi^2}}.
\end{align}
This gives the four twist-3 GPDs $E_2^\mathcal{M}$, $F_2^\mathcal{M}$, $G_2^\mathcal{M}$ and $H_2^\mathcal{M}$ from their correlators. Note that $F_2^\mathcal{M}$ corresponds to the vector projection while $H_2^\mathcal{M}$ corresponds to the tensor projection,
\begin{align}
    F^{[\mathbf{1}]}                & = \int d^2\bm{k}_\perp \, \mathcal{N}_3 \Big[ 2m_1 x S + i (\bm{A} \times \bm{V}) \Big],\label{eq:gpd_E2_dec}                                                    \\
    \bm{F}^{[\gamma^\perp]}         & = \int d^2\bm{k}_\perp \, \mathcal{N}_3 \Big[ 2 i m_1 \xi (\hat{\bm{z}} \times \bm{V}) + S \bm{B} - i P (\hat{\bm{z}} \times \bm{A}) \Big],\label{eq:gpd_F2_dec} \\
    F^{[i\sigma^{12}\gamma_5]}      & = \int d^2\bm{k}_\perp \, \mathcal{N}_3 \Big[ 2 i m_1 \xi S - (\bm{B} \times \bm{V}) \Big], \label{eq:gpd_H2_dec}                                                \\
    (F^{[i\sigma^{21}\gamma_5]}     & = - F^{[i\sigma^{12}\gamma_5]}),\nonumber                                                                                                                        \\
    \bm{F}^{[\gamma^\perp\gamma^5]} & = \int d^2\bm{k}_\perp \, \mathcal{N}_3 \Big[ 2 m_1 x \bm{V} + P \bm{B} - i S (\hat{\bm{z}} \times \bm{A}) \Big].
    \label{eq:gpd_G2_dec}
\end{align}

\subsection{Twist-4 Correlators}
The twist-4 correlators feature highly coupled quadratic momentum structures. The tensor $\mathcal{T}$ emerges naturally from the interference of angular phases with the vector amplitude. We define the kinematic coefficient,
\begin{align}
    \mathcal{N}_4 = \frac{1}{64 \pi^3 x (P^+)^2 \sqrt{x^2-\xi^2}}.
\end{align}
This gives the two twist-4 GPDs $F_3^\mathcal{M}$ and $H_3^\mathcal{M}$ from their correlators,
\begin{align}
    F^{[\gamma^-]}                      & = \int d^2\bm{k}_\perp \, \mathcal{N}_4 \Big[ (4\bm{k}_\perp^2 + 4m_1^2 - \bm{\Delta}_\perp^2) S - 4 i m_1 (\bm{\Delta}_\perp \times \bm{V}) + 4 i P (\bm{k}_\perp \times \bm{\Delta}_\perp) \Big],\label{eq:gpd_F3_dec}              \\
    \bm{F}^{[i\sigma^{\perp-}\gamma_5]} & = \int d^2\bm{k}_\perp \, \mathcal{N}_4 \Big[ 4 m_1^2 \bm{V} - 4 \mathcal{T}(\bm{k}_\perp, \bm{V}) + \mathcal{T}(\bm{\Delta}_\perp, \bm{V}) + 4 m_1 \big( 2 P \bm{k}_\perp - i S (\hat{\bm{z}} \times \bm{\Delta}_\perp) \big) \Big].
    \label{eq:gpd_H3_dec}
\end{align}
\section{Fourier transforms of GPDs}

The 3D tomographic picture of the meson is revealed by Fourier transforming the GPDs from momentum space to coordinate space~\cite{Diehl:2002he, Burkardt:2000za, Burkardt:2003je}.

\subsection{Impact parameter dependent parton distribution functions}

In the zero-skewness limit ($\xi = 0$), the invariant momentum transfer is purely transverse, $t = -\bm{\Delta}_\perp^2$. The two-dimensional Fourier transform of a GPD with respect to the transverse momentum transfer $\bm{\Delta}_\perp$ yields the IPDPDF. For a GPD $F^\mathcal{M}$, the IPDPDF is defined as
\begin{align}
    q^\mathcal{M}(x, \bm{b}_\perp) & = \int \frac{d^2 \bm{\Delta}_\perp}{(2\pi)^2} e^{-i \bm{\Delta}_\perp \cdot \bm{b}_\perp} F^\mathcal{M}(x, 0, -\bm{\Delta}_\perp^2),
\end{align}
where $\bm{b}_\perp$ is the transverse impact parameter, representing the distance between the active quark and the transverse center of momentum of the meson. This distribution allows us to visualize the spatial distribution of partons in the transverse plane. Integrating over the transverse coordinate space recovers the standard collinear PDF, $f(x) = \int d^2 \bm{b}_\perp \, q^\mathcal{M}(x, \bm{b}_\perp)$.

\subsection{Longitudinal position space distributions}

To explore the distribution of partons in the longitudinal direction, we perform a Fourier transform of the GPDs with respect to the skewness parameter $\xi$ for a fixed value of the invariant momentum transfer squared $t$. The conjugate variable to the skewness is denoted by $\sigma$, which serves as a measure of the longitudinal distance. The distribution in longitudinal position space is defined as
\begin{align}
    S^\mathcal{M}(x, \sigma, t) & = \int \frac{d\xi}{2\pi} e^{-i \xi \sigma} F^\mathcal{M}(x, \xi, t),
\end{align}
where $F^\mathcal{M}(x, \xi, t)$ represents a GPD. By analyzing these distributions at fixed invariant momentum transfer ($t < 0$), we can investigate the longitudinal localization of quarks carrying a specific longitudinal momentum fraction $x$.

\section{Results and discussion}

Before proceeding, it is necessary to clarify the generic meson notation adopted in this work. Physically, the pion exists in three distinct charge states ($\pi^+, \pi^-, \pi^0$), and the kaon exists in four ($K^+, K^-, K^0, \bar{K}^0$). However, within the framework of our LFQM, we operate in the strict isospin symmetry limit where the up and down constituent quark (antiquark) masses $m_u$ and $m_d$ respectively, are treated as degenerate ($m_u = m_d$). Furthermore, since we do not explicitly incorporate electromagnetic interactions, neglecting the subtle macroscopic mass splittings between the charged and neutral states of these mesons. Consequently, the internal partonic dynamics and the resulting GPDs evaluated in this model are dynamically equivalent across all states within a given meson multiplet. For this reason, throughout this paper, we consistently employ the generic notations $\pi$ and $K$ to represent any physical charge state of the pion and kaon, respectively. When isolating the contributions of individual valence constituents, we formalize the notation to $\mathcal{M}_q$, where $\mathcal{M}$ denotes the meson and $q$ specifies the active quark flavor. Accordingly, in the numerical evaluations that follow, we explicitly present the calculated distributions for the up quark in the pion ($\pi_u$) as the representative symmetric case, alongside both the up ($K_u$) and strange antiquark ($K_{\bar{s}}$) in the kaon to explicitly illustrate the dynamical consequences of $\mathrm{SU}(3)$ flavor symmetry breaking.

The complete set of pion and kaon GPDs, defined in Eqs.~\eqref{eq:gpd_F1}--\eqref{eq:gpd_H3} and Eqs.~\eqref{eq:gpd_F1_dec}--\eqref{eq:gpd_H3_dec}, are numerically evaluated using the LFQM described in the previous section. The full set of numerical parameters used for this phenomenological model is summarized in Table~\ref{tab:parameters}. The constituent quark masses, meson masses, and oscillator parameters are chosen to be consistent with established literature~\cite{Luan:2024dvc}. These phenomenological parameters were derived from the global analysis by Qian and Ma \cite{Qian:2008px}, fitting a wide array of low-energy observables. We note that the large constituent mass ratio $m_s/m_u \approx 2.8$ adopted in our LFQM parameterization inherently leads to a pronounced flavor asymmetry in the kaon GPDs, which we will discuss in the subsequent subsections. The normalization constants are uniquely fixed by enforcing the probability normalization condition of the lowest-lying Fock state. Because these GPDs are evaluated strictly within the DGLAP kinematic domain, results are presented for $x > \xi$. Computations are performed at the initial model scale, providing the baseline boundary conditions required for subsequent DGLAP evolution to higher experimental scales.

\begin{table}[htpb]
    \caption{\label{tab:parameters} Model parameters for the pion ($\pi$) and kaon ($K$). The active and spectator constituent quark masses ($m_u, m_s$), the macroscopic meson masses ($M_{\pi, K}$), and the harmonic oscillator parameters ($\beta_{\pi, K}$) are given in GeV. The normalization constants $A^\mathcal{M}$ (in GeV$^{-1}$) are constrained by the probability normalization of the wavefunctions.}
    \begin{ruledtabular}
        \begin{tabular}{lc}
            Parameter                                                  & Value     \\
            \hline
            $u/d$-quark/antiquark mass $m_u$                           & $0.200$   \\
            $s$-quark/antiquark mass $m_s$                             & $0.556$   \\
            Mass of the pion $M_\pi$                                   & $0.140$   \\
            Mass of the kaon $M_K$                                     & $0.493$   \\
            Oscillator parameter of pion $\beta_\pi$                   & $0.410$   \\
            Oscillator parameter of kaon $\beta_K$                     & $0.405$   \\
            Normalization constant ($m_1 = m_u$, $m_2 = m_u$), $A^\pi$ & $44.2357$ \\
            Normalization constant ($m_1 = m_u$, $m_2 = m_s$), $A^K$   & $74.0333$ \\
        \end{tabular}
    \end{ruledtabular}
\end{table}

\subsection{Numerical results for GPDs at fixed $\xi$}
\begin{figure}[!ht]
    \centering
    \begin{minipage}{\linewidth}
        \includegraphics[width=\linewidth]{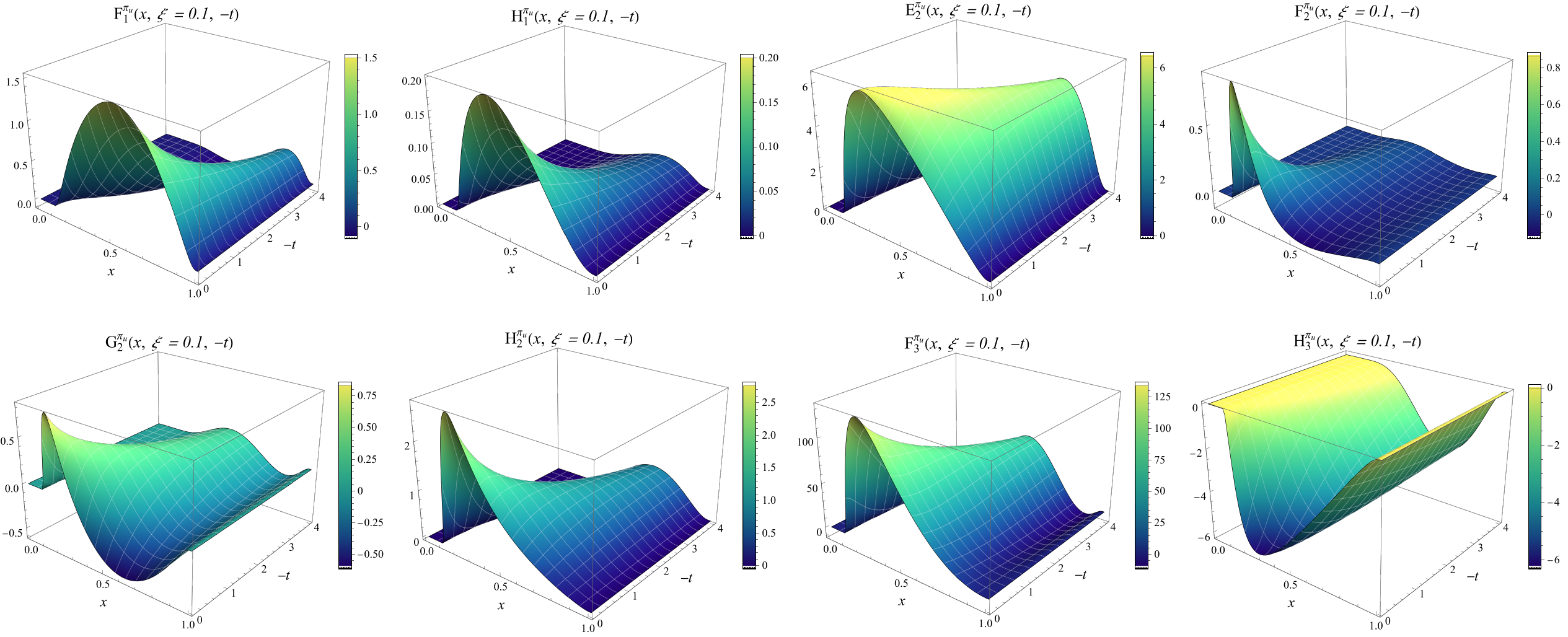}
    \end{minipage}

    \vspace{0.5cm}

    \begin{minipage}{\linewidth}
        \includegraphics[width=\linewidth]{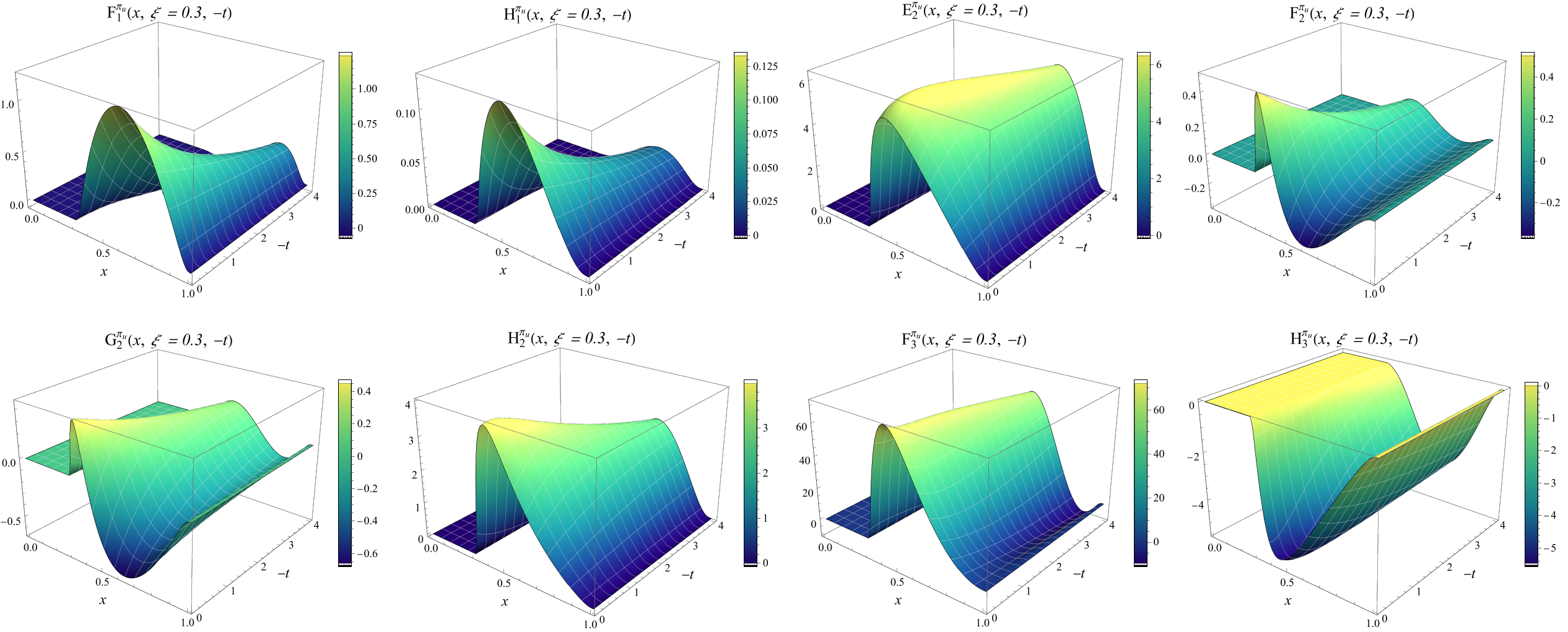}
    \end{minipage}

    \caption{The eight GPDs of the pion ($u$ quark) as functions of the longitudinal momentum fraction $x$ and the invariant momentum transfer squared $-t$, evaluated at fixed skewness $\xi = 0.1$ (top two rows) and $\xi = 0.3$ (bottom two rows). The set includes twist-2 ($F_1^{\pi_u}, H_1^{\pi_u}$), twist-3 ($E_2^{\pi_u}, F_2^{\pi_u}, G_2^{\pi_u}, H_2^{\pi_u}$), and twist-4 ($F_3^{\pi_u}, H_3^{\pi_u}$) distributions.}
    \label{fig:pion_gpds_xi}
\end{figure}

The 3D surface plots in Fig.~\ref{fig:pion_gpds_xi} illustrate the GPDs of pion, which shows the interplay between $x$ and $-t$ at fixed skewness values $\xi=0.1$ (top two rows) and $\xi=0.3$ (last two rows). Universally, the GPD amplitudes decay monotonically with increasing $-t$, consistent with the expected behavior of hadronic form factors reflecting the finite transverse spatial extent of the meson. At $\xi = 0.1$, shown in the first two rows of Fig.~\ref{fig:pion_gpds_xi}, the twist-2 distributions $F_1^{\pi_u}$ and $H_1^{\pi_u}$ exhibit pronounced symmetric peaks with respect to $x$ at low momentum transfer, reaching maximum amplitudes of 1.53 at $x=0.53$ and 0.20 at $x=0.36$, respectively. Moving to higher twists, $E_2^{\pi_u}$ displays a gradual decline with respect to $-t$ from its peak value of 6.44. In contrast, $F_2^{\pi_u}$ drops sharply along the $-t$ axis and develops negative probability domains at higher $x$, falling to a minimum of $-0.12$. The chiral-odd $G_2^{\pi_u}$ is uniquely characterized by distinct positive and negative peaks, 0.83 at $x=0.15$ and $-0.59$ at $x=0.62$ respectively, while $H_2^{\pi_u}$ presents a positive peak of 2.76 near $x=0.19$. The twist-4 distributions exhibit significant amplitude scaling, as seen from the plots, $F_3^{\pi_u}$ reaches 133.62 at low $x$, particularly at $x=0.23$, while $H_3^{\pi_u}$ forms an entirely negative distribution with a peak value of $-6.21$ at $x=0.28$.

Increasing the skewness to $\xi = 0.3$ (presented in the lower two rows of Fig.~\ref{fig:pion_gpds_xi}) generally shifts the distribution to higher $x$ and also suppresses the overall magnitude of the distributions. For instance, the $F_1^{\pi_u}$ peak drops to 1.24 while shifting to a higher $x$ value of 0.61, and this qualitative behavior is common for other distributions $H_1^{\pi_u}$, $E_2^{\pi_u}$, $F_2^{\pi_u}$, $G_2^{\pi_u}$, $F_3^{\pi_u}$ and $H_3^{\pi_u}$. A notable exception to this suppression pattern can be observed for $H_2^{\pi_u}$, which scales up to 3.99. The kinematic shift also alters the topology of the higher-twist functions. The peak of $E_2^{\pi_u}$ (6.31) migrates from strictly $-t \approx 0$ to $-t = 1.04 \text{ GeV}^2$. Furthermore, the negative domains in $F_2^{\pi_u}$ and $G_2^{\pi_u}$ become more pronounced, dropping to $-0.35$ and $-0.66$, respectively, as the larger longitudinal momentum transfer seen through increased $\xi$ forces greater correlations between the initial and final states. In twist-4, the $H_3^{\pi_u}$ scaled down in the negative direction to a peak value $-5.4$.

\begin{figure}[!ht]
    \centering
    \begin{minipage}{\linewidth}
        \includegraphics[width=\linewidth]{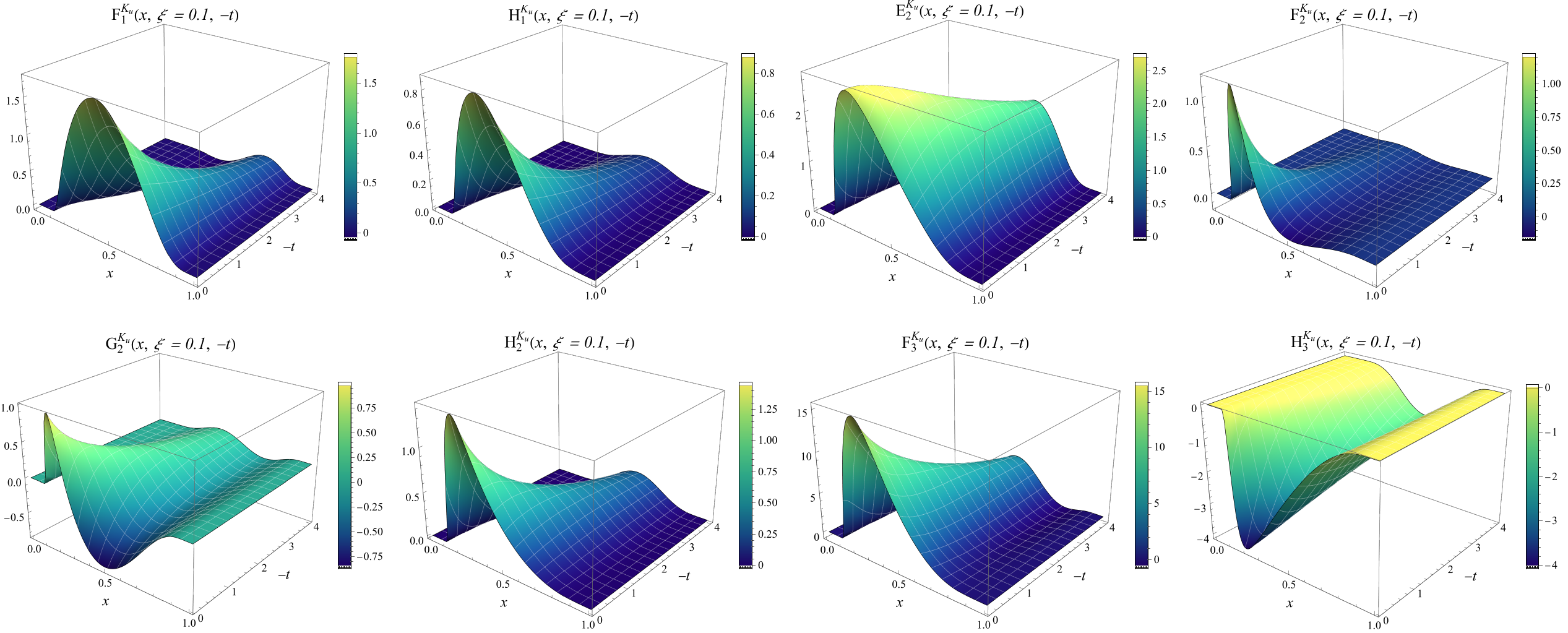}
    \end{minipage}

    \vspace{0.5cm}

    \begin{minipage}{\linewidth}
        \includegraphics[width=\linewidth]{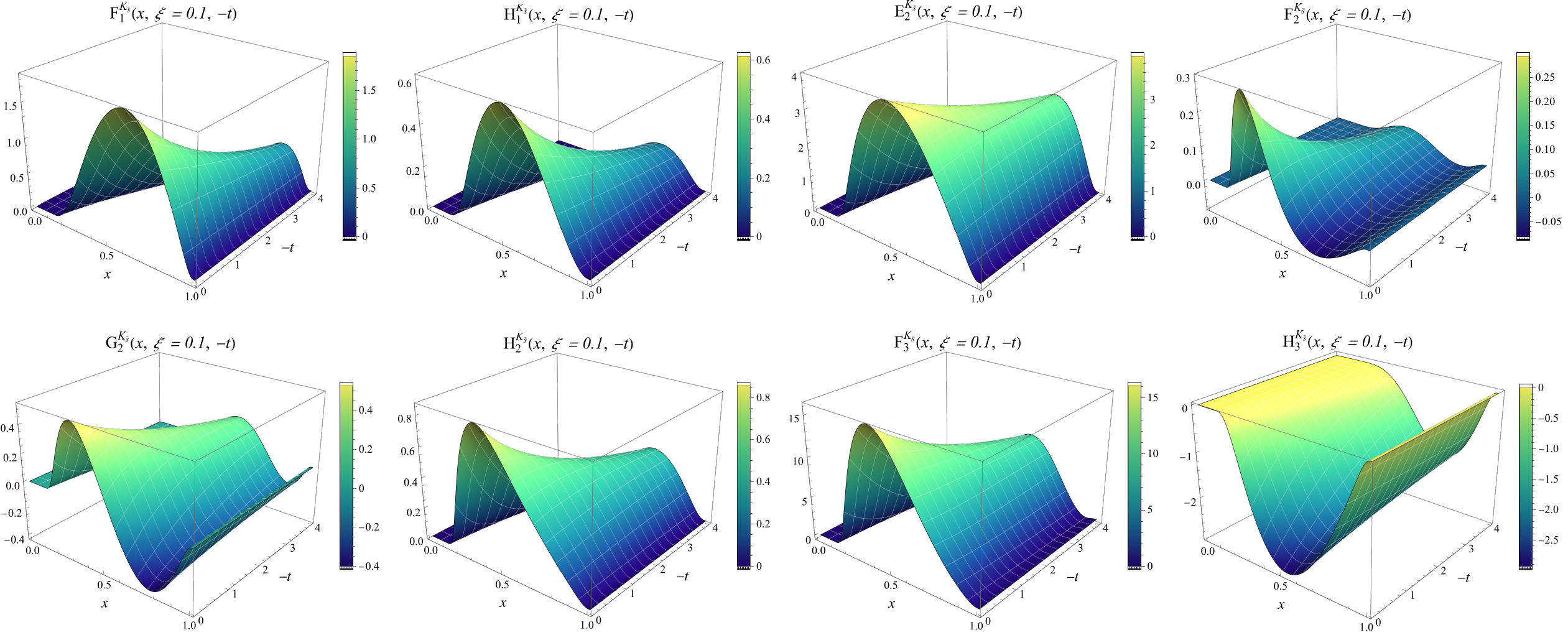}
    \end{minipage}

    \caption{The eight GPDs of the kaon as a function of longitudinal momentum fraction $x$ and invariant momentum transfered squared $-t$, evaluated at skewness parameter $\xi = 0.1$, comparing the distributions of the lighter $u$ quark (top two rows) and the heavier $\bar{s}$ quark (bottom two rows).}
    \label{fig:kaon_gpds_xi}
\end{figure}

Figure~\ref{fig:kaon_gpds_xi} shows the corresponding fixed-$\xi$ GPDs for the kaon, separating the contributions of the $u$ and $\bar{s}$ quarks in the top two rows and lower two rows respectively, to highlight explicit $\mathrm{SU}(3)$ flavor symmetry breaking. While the global decay with respect to $-t$ mirrors the pion, the mass asymmetry between the constituent quarks modifies their longitudinal momentum distributions. The heavier $\bar{s}$ quark naturally carries a disproportionately larger fraction of the kaon's momentum compared to the lighter $u$ quark within the kaon, which explicitly shifts the peaks of all $\bar{s}$ GPDs toward higher $x$ values relative to the $u$ quark. For instance, at $\xi=0.1$, the twist-2 $F_1^{K_{\bar{s}}}$ peaks at $x=0.61$ with an amplitude of 1.85, whereas the $u$ quark distribution $F_1^{K_u}$ peaks earlier at $x=0.43$ with an amplitude of 1.76. The twist-2 $H_1$ behaves similarly, with $H_1^{K_{\bar{s}}}$ peaking at 0.61 and $H_1^{K_u}$ at 0.88. Among the twist-3 distributions, $E_2^{K_{\bar{s}}}$ and $H_2^{K_{\bar{s}}}$ achieve maximum amplitudes of 3.96 and 0.86 respectively, while their $u$ quark counterparts peak at 2.70 and 1.44. The chiral distributions $F_2$ and $G_2$ develop both positive peaks and negative troughs; for example, $F_2^{K_{\bar{s}}}$ ranges from 0.29 to $-0.08$, whereas $F_2^{K_u}$ exhibits a stronger positive peak of 1.20 and a trough of $-0.15$. Similar kinematic shifts persist through the higher twists. The twist-4 $F_3^{K_{\bar{s}}}$ distribution reaches a peak of 16.13 at $x=0.37$, while $F_3^{K_u}$ peaks at 15.48 earlier at $x=0.22$. The predominately negative $H_3$ distribution shows a minimum of $-2.93$ for $H_3^{K_{\bar{s}}}$ compared to $-4.07$ for $H_3^{K_u}$.

\subsection{Numerical results for GPDs at fixed $-t$}

\begin{figure}[!ht]
    \centering
    \begin{minipage}{\linewidth}
        \includegraphics[width=\linewidth]{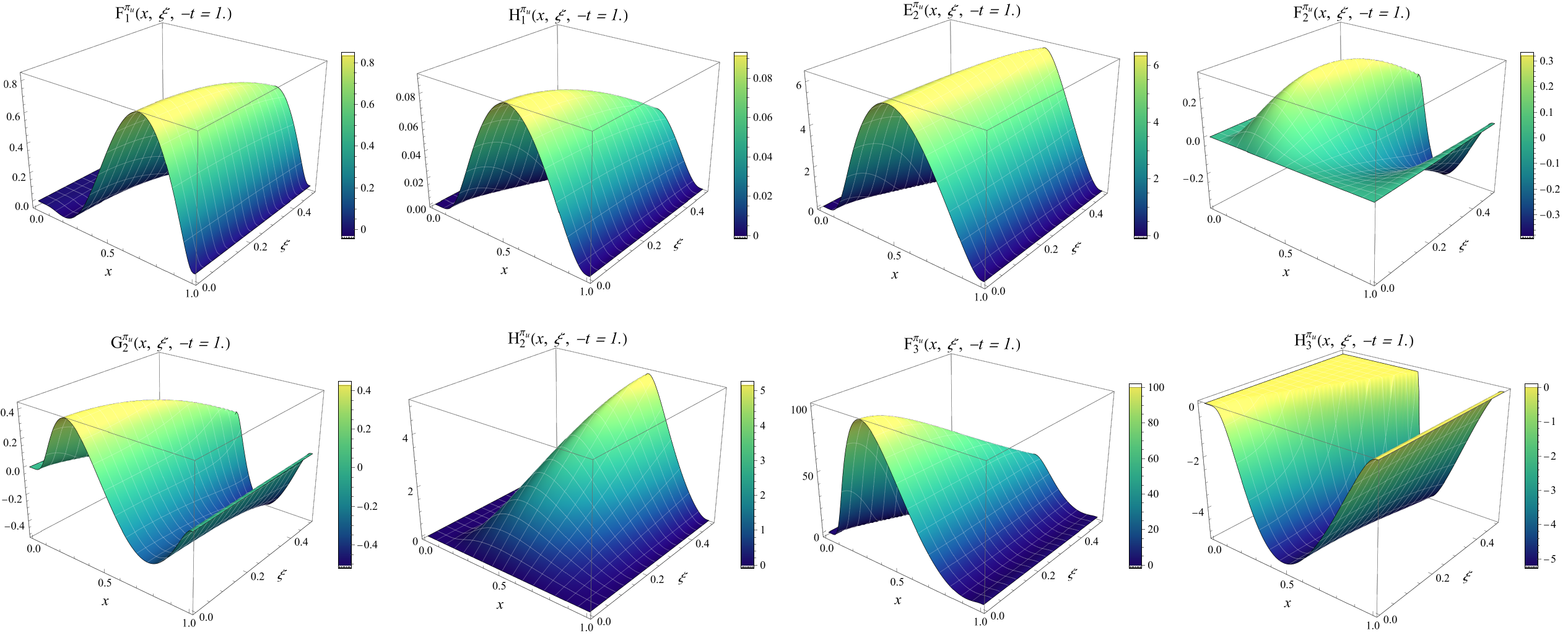}
    \end{minipage}

    \vspace{0.5cm}

    \begin{minipage}{\linewidth}
        \includegraphics[width=\linewidth]{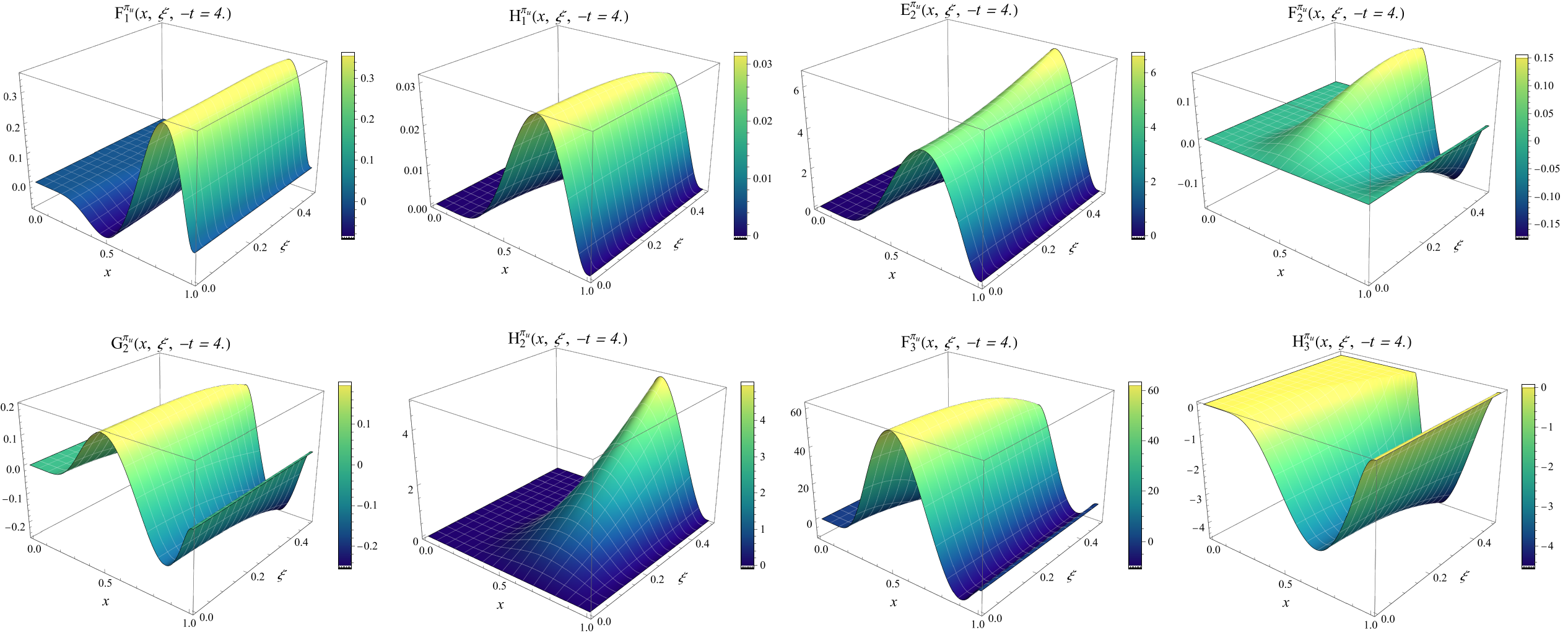}
    \end{minipage}

    \caption{The GPDs of the pion $u$ quark as functions of longitudinal momentum fraction $x$ and skewness parameter $\xi$, evaluated at a fixed momentum transfer of $-t = 1.0 \text{ GeV}^2$ shown in the top two rows and $-t = 4.0 \text{ GeV}^2$ shown in the bottom two rows.}
    \label{fig:pion_gpds_t}
\end{figure}

The dependence of the GPDs on the skewness parameter $\xi$ and longitudinal momentum fraction $x$ is explored in Fig.~\ref{fig:pion_gpds_t} at fixed momentum transfers of $-t = 1.0 \text{ GeV}^2$ (top two rows) and $4.0 \text{ GeV}^2$ (last two rows) for pion. Among the complete set of distributions, the behavior of the twist-3 chiral-even GPD $F_2^{\pi_u}$ and the chiral-odd GPD $H_2^{\pi_u}$ is fundamentally governed by the hermiticity constraints of their underlying generalized transverse momentum dependent distributions (GTMDs)~\cite{Meissner:2008ay}. When integrated over transverse momentum to project onto the GPD limit, hermiticity explicitly requires $F_2^{\pi_u}$ and $H_2^{\pi_u}$ to be odd functions of $\xi$. A direct mathematical consequence of this odd symmetry is that both distributions must strictly vanish in the forward limit, yielding $F_2^{\pi_u}(x,0,t) = 0$ and $H_2^{\pi_u}(x,0,t) = 0$. Furthermore, because establishing structural relations between GPDs and standard TMDs inherently relies on evaluating the system at $\xi = 0$, this vanishing behavior precludes $F_2^{\pi_u}$ and $H_2^{\pi_u}$ from having direct TMD analogs. This rigorous constraint is clearly visible in Fig.~\ref{fig:pion_gpds_t}, where both these distributions fall to zero at $\xi=0$ and exhibit clear odd symmetry along the skewness axis.

The GPDs of pion at $-t = 1.0 \text{ GeV}^2$ can be seen plotted in the first two rows of Fig.~\ref{fig:pion_gpds_t}. The twist-2 GPDs $F_1^{\pi_u}$ and $H_1^{\pi_u}$ peak near the forward limit ($\xi \to 0$) with maximum amplitudes of 0.83 at $x=0.70$ and 0.09 at $x=0.54$, respectively. Both distributions are heavily localized in the valence-dominated high-$x$ region and exhibit a slow decline as skewness increases. The twist-3 $E_2^{\pi_u}$ shows similar stability with respect to $\xi$, maintaining a peak amplitude of 6.35 broadly centered around $x=0.56$. The $G_2^{\pi_u}$ distribution retains its dual-node topology along the $x$ axis regardless of skewness, characterized by a positive peak of 0.43 at lower momentum fractions ($x=0.27$) and a distinct negative trough at $-0.50$ in the higher-$x$ regime ($x=0.73$). For the odd-symmetry distributions, $F_2^{\pi_u}$ and $H_2^{\pi_u}$ naturally develop substantial probability densities only at higher skewness, reaching peaks of 0.32 (at $x=0.36$) and 5.16 (at $x=0.60$), respectively. Moving to the twist-4 sector, $F_3^{\pi_u}$ presents a prominent positive peak of 100.46 heavily localized in the lower-$x$ region ($x=0.29$) at $\xi=0$. Conversely, $H_3^{\pi_u}$ maintains an entirely negative amplitude, reaching $-5.18$ at $x=0.55$, that spreads broadly across the intermediate momentum fractions.

Increasing the momentum transfer to $-t = 4.0 \text{ GeV}^2$, shown at the lower two rows of Fig.~\ref{fig:pion_gpds_t}, introduces a universal amplitude suppression across all distributions while preserving their underlying topological symmetries. This systematic decay is a consequence of structural form factor scaling. Higher momentum transfers probe increasingly compact spatial configurations within the meson. For instance, the primary peak of the twist-2 unpolarized distribution $F_1^{\pi_u}$ drops from 0.83 at $1.0 \text{ GeV}^2$ down to 0.35 at $4.0 \text{ GeV}^2$, while its maximum shifts to $x=0.82$. A similar proportional suppression is observed in the GPD $H_1^{\pi_u}$, whose peak falls to 0.03. Crucially, the strict hermiticity constraints dictating the behavior of the twist-3 functions remain fully intact. Both $F_2^{\pi_u}$ and $H_2^{\pi_u}$ continue to strictly vanish along the forward $\xi = 0$ axis. In the twist-4 sector, the $F_3^{\pi_u}$ peak is suppressed from 100.46 down to 62.07. Despite this amplitude reduction across all twists, the relative functional shapes and specific $x$-localizations are cleanly preserved across the $(x, \xi)$ kinematic plane.

\begin{figure}[!ht]
    \centering
    \begin{minipage}{\linewidth}
        \includegraphics[width=\linewidth]{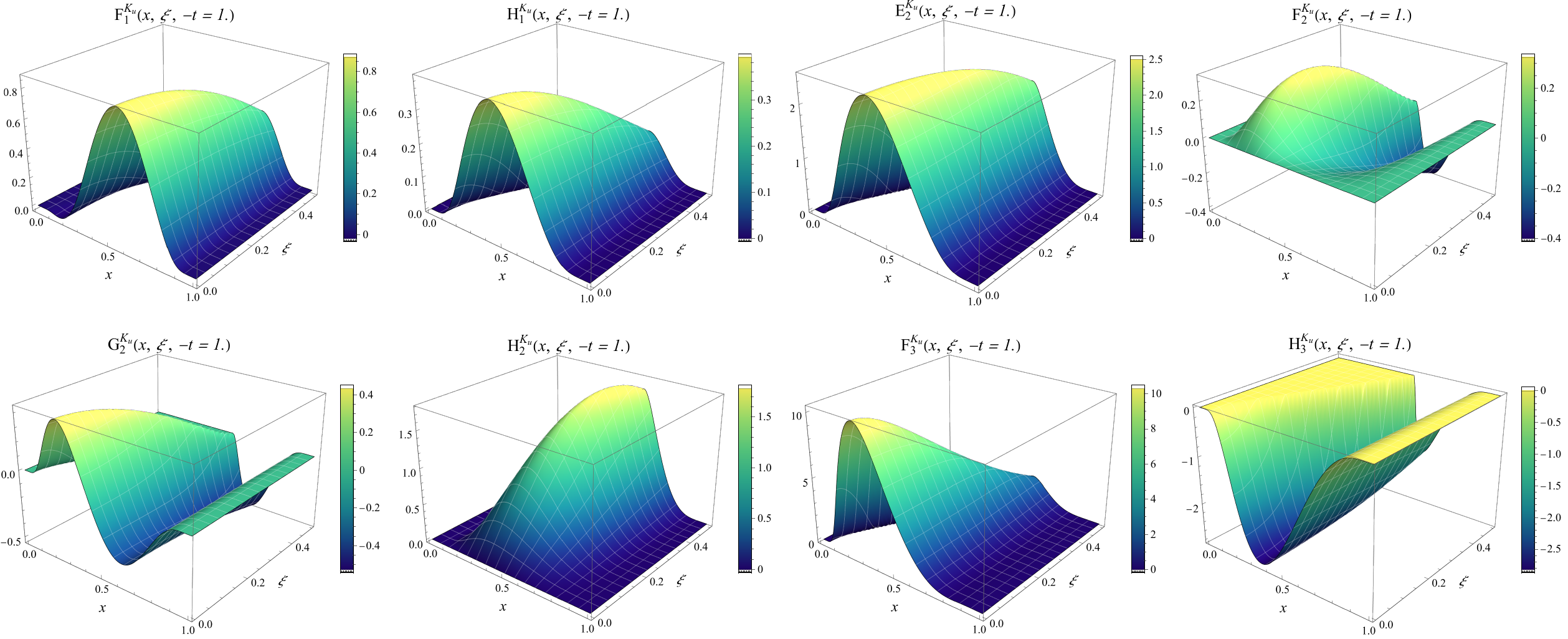}
    \end{minipage}

    \vspace{0.5cm}

    \begin{minipage}{\linewidth}
        \includegraphics[width=\linewidth]{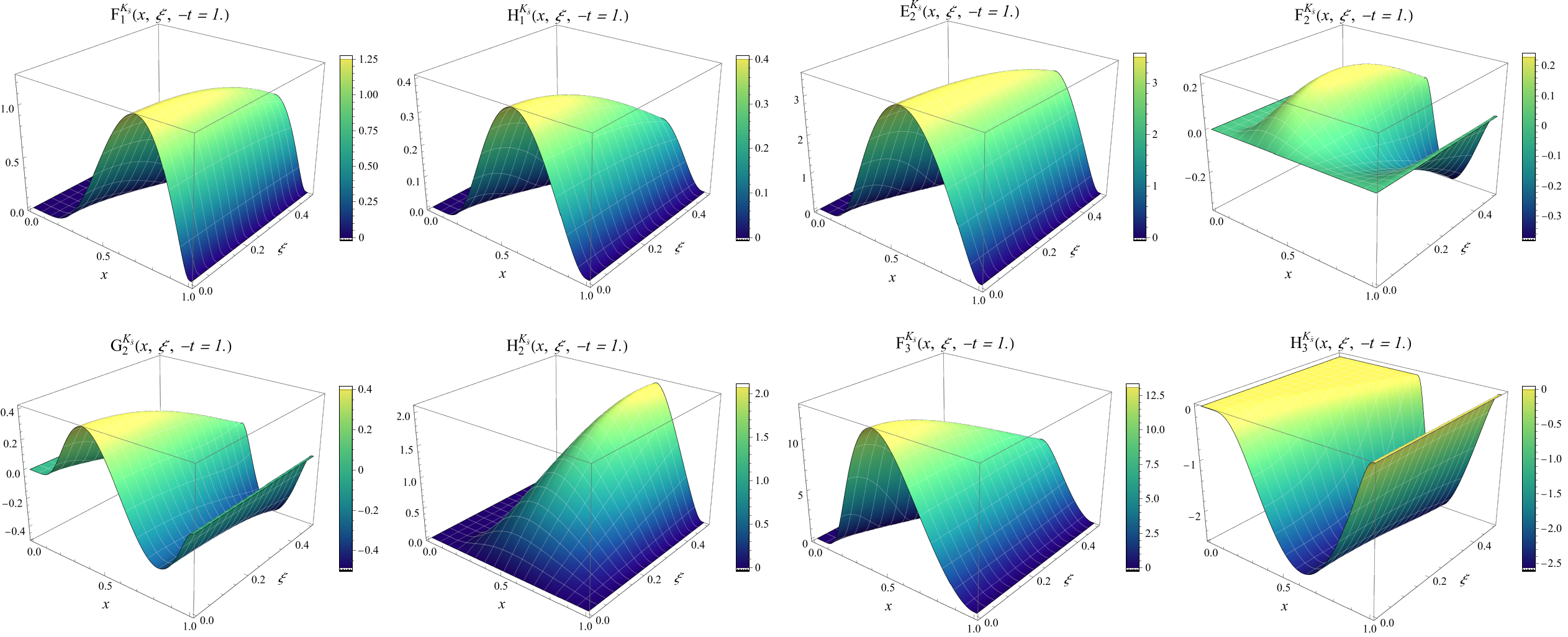}
    \end{minipage}

    \caption{The kaon GPDs as functions of longitudinal momentum fraction $x$ and skewness parameter $\xi$ at fixed momentum transfer $-t = 1.0 \text{ GeV}^2$, comparing the $u$ quark (top two rows) and $\bar{s}$ quark (bottom two rows).}
    \label{fig:kaon_gpds_t}
\end{figure}

Figure~\ref{fig:kaon_gpds_t} depicts the fixed-t GPDs for the kaon $u$ quark in the first two rows and the $\bar{s}$ quark in the last two rows, at $-t = 1.0 \text{ GeV}^2$. The topological behavior across the $\xi$ domain closely resembles that of the pion, but the flavor asymmetry again dictates the amplitude scales. At this specific momentum transfer, the kaon's $\bar{s}$ quark distributions exhibit systematically higher peak amplitudes at low $\xi$ compared to the corresponding $u$ quark distributions. Looking at the leading twist distributions, $F_1^{K_{\bar{s}}}$ peaks at 1.25, while $F_1^{K_u}$ peaks at 0.87, and similarly $H_1^{K_{\bar{s}}}$ reaches 0.40 compared to 0.39 for $H_1^{K_u}$. This scaling trend persists comprehensively across the higher twists. The twist-3 $E_2^{K_{\bar{s}}}$ peaks scale to 3.51 for $\bar{s}$ versus $E_2^{K_{u}}$ up to 2.51 for $u$, and the $H_2$ distribution reaches 2.07 for $\bar{s}$ versus 1.78 for $u$. The odd-symmetry distributions $F_2$ and $G_2$ also exhibit distinct domains; $F_2^{K_{\bar{s}}}$ displays a peak of 0.23 and a trough of $-0.37$, while $F_2^{K_u}$ reaches 0.32 and $-0.41$, with $G_2$ showing parallel behavior. The twist-4 dynamics follow identically, with $F_3^{K_{\bar{s}}}$ peaking at 13.09 for $\bar{s}$ versus $F_3^{K_{u}}$ at 10.29 for $u$, and the negative $H_3^{K_{\bar{s}}}$ distribution dropping to $-2.56$ for $\bar{s}$ versus $H_3^{K_{u}}$ to $-2.82$ for $u$. As observed in the fixed-$\xi$ analysis, all $\bar{s}$ distributions remain systematically shifted toward higher $x$ across the entire skewness domain, reflecting the persistent kinematic dominance of the heavier strange constituent.

\subsection{Numerical results for IPDPDFs}

In this subsection, we analyze the spatial structure of the pseudoscalar mesons by examining their IPDPDFs in the transverse plane. As discussed in the previous section, the Fourier transform of the GPDs at $\xi=0$ yields these 2D spatial densities, $q^\mathcal{M}(x,\mathbf{b_\perp})$, which describe the probability density of finding an active quark at a transverse distance $b_\perp$ carrying a longitudinal momentum fraction $x$. From the eight GPDs discussed earlier, only six IPDPDFs are possible as $F_2^{\mathcal{M}}$ and $H_2^{\mathcal{M}}$ strictly vanish at $\xi=0$ where IPDPDFs are defined.

\begin{figure}[!ht]
    \centering
    \begin{minipage}{\linewidth}
        \includegraphics[width=\linewidth]{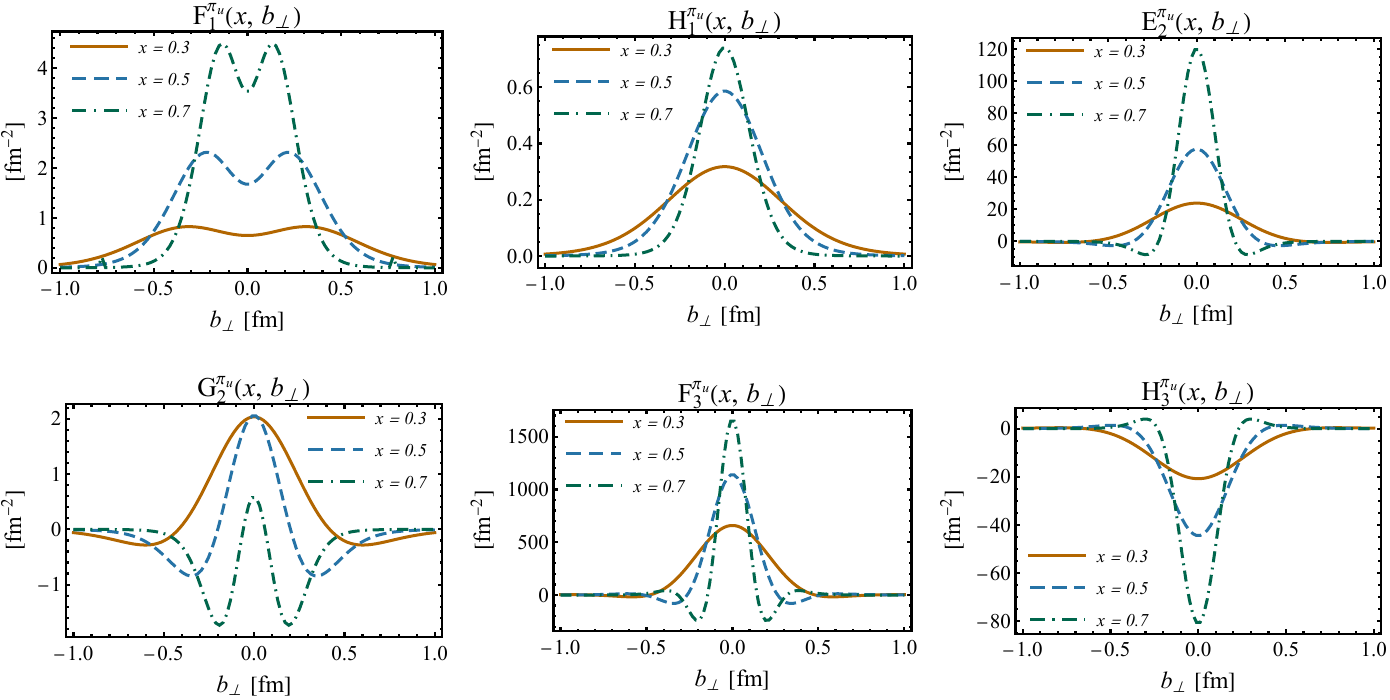}
    \end{minipage}

    \caption{The six IPDPDFs (in fm$^{-2}$) of pion at as a function of transverse spatial distance $b_\perp$ (in fm) for three longitudinal momentum fraction values $x$ at 0.3, 0.5, and 0.7.}
    \label{fig:pion_ipdpdf}
\end{figure}

In Fig.~\ref{fig:pion_ipdpdf}, all possible IPDPDFs (in fm$^{-2}$) of the pion $u$ quark are plotted as a function of transverse spatial distance $b_\perp$ (in fm)  for different longitudinal momentum fraction values $x$. Across all these distributions, a universal scaling behavior is immediately evident. As the longitudinal momentum fraction increases from $x = 0.3$ to $x = 0.7$, the transverse distributions systematically narrow. This physical contraction reflects the correlation between longitudinal and transverse degrees of freedom. At high $x$, the active quark carries the majority of the pion's momentum, forcing the transverse separation between the active quark and the spectator to shrink. Consequently, the probability density becomes highly localized near the transverse center of mass ($b_\perp\to0$).

At lower momentum fraction, $x=0.3$, $F_1^{\pi_u}(x, b_\perp)$ is broad and centrally peaked at 0.83. However, as $x$ increases, the overall amplitude scales to 4.49 at $x=0.7$ and a pronounced bimodal structure emerges. A central dip forms at $b_\perp = 0$, pushing the maximum probability density outward into a ring-like structure in the transverse plane. It is important to note that this bimodal structure of $F_1^{\pi_u}$ when projected to the impact parameter space is highly sensitive to our specific parameterization of the LFWF. For the symmetric pion, the absence of the explicit mass-splitting term in the BHL exponent preserves this bimodal topology. In contrast to the unpolarized case, $H_1^{\pi_u}(x, b_\perp)$ maintains a strict, unimodal Gaussian-like profile across all calculated $x$ values. The peak resides permanently at the center and scales steadily in amplitude, from 0.32 at $x=0.3$ to 0.74 at $x=0.7$, while the tails compress inward. The twist-3 $E_2^{\pi_u} (x, b_\perp)$ features a prominent central peak that scales with $x$, reaching an amplitude of 119.86 at $x = 0.7$. Unlike the twist-2 functions, $E_2^{\pi_u}$ develops negative probability domains, with negative peaks at $-8.31$, appearing symmetrically around the central core. The $G_2^{\pi_u} (x, b_\perp)$ distribution displays a distinct topological transition as a function of $x$. At $x = 0.3$, it presents as a broad, purely positive central peak. By $x = 0.7$, negative troughs at $-1.73$ dominate the intermediate $b_\perp$ regions. The amplitude of the twist-4 $F_3^{\pi_u} (x, b_\perp)$ shows a significant scale amplification, reaching a peak value of 1647.40 at $x = 0.7$. $H_3^{\pi_u} (x, b_\perp)$ shows a predominately negative distribution, with the minimum decreasing from $-20.76$ to $-80.44$ as $x$ increases from 0.3 to 0.7.

\begin{figure}[!ht]
    \centering
    \begin{minipage}{\linewidth}
        \includegraphics[width=\linewidth]{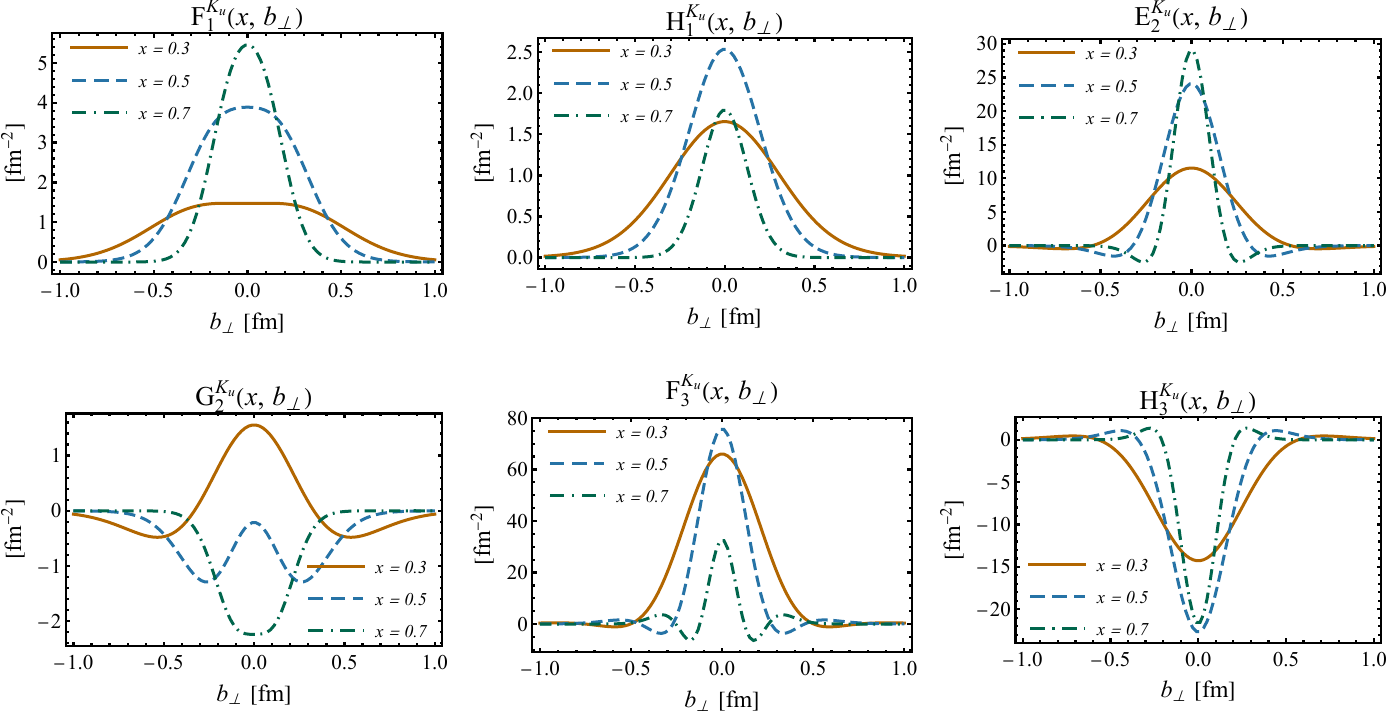}
    \end{minipage}

    \vspace{0.5cm}

    \begin{minipage}{\linewidth}
        \includegraphics[width=\linewidth]{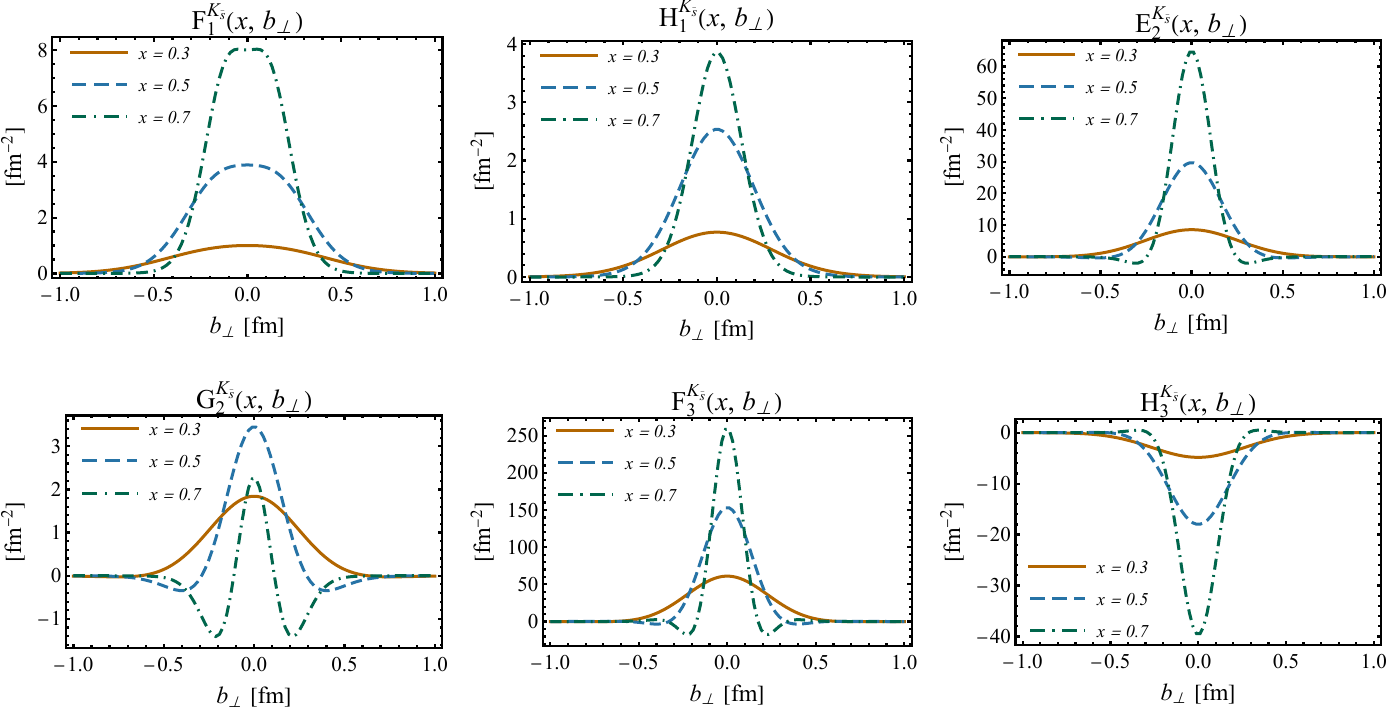}
    \end{minipage}

    \caption{The first six and the next six plots show the IPDPDFs (in fm$^{-2}$) of kaon $u$ quark and $\bar{s}$ quarks, respectively, as a function of transverse spatial distance $b_\perp$ (in fm), at three longitudinal momentum fraction values $x$ at 0.3, 0.5, and 0.7.}
    \label{fig:kaon_ipdpdf}
\end{figure}

The IPDPDFs (in fm$^{-2}$) for both the $u$ quark and the $\bar{s}$ quark in the kaon are presented in Fig.~\ref{fig:kaon_ipdpdf} in the first two rows and the last two rows, respectively. A notable difference from the pion distributions is the explicit manifestation of $\mathrm{SU}(3)$ flavor symmetry breaking. For the lighter $u$ quark, the spatial distributions exhibit non-monotonic amplitude scaling at high $x$. For example, in $H_1^{K_u}(x, b_\perp)$ and $F_3^{K_u}(x, b_\perp)$, the central probability peak reaches a maximum at $x=0.5$ (2.53 and 75.55, respectively) before dropping at $x=0.7$ (to 1.79 and 32.41). The twist-3 $E_2^{K_u}$ peaks from 11.52 at $x=0.3$ to 28.94 at $x=0.7$, while the negative $H_3^{K_u}$ minimum deepens from $-14.26$ to $-21.55$. Furthermore, $G_2^{K_u}(x, b_\perp)$ undergoes a distinct topological inversion, where the central core transitions from a positive peak at $x=0.3$ into a negative trough value of $-2.24$ at $x=0.7$.

Conversely, the distributions for the heavier $\bar{s}$ quark, $F_1^{K_{\bar{s}}}$, $H_1^{K_{\bar{s}}}$, and $E_2^{K_{\bar{s}}}$, display amplitude amplification up to $x=0.7$, reaching central peaks of 8.04, 3.85, and 64.55, respectively. The $G_2^{K_{\bar{s}}}$ distribution also exhibits growth, increasing from a peak of 1.84 at $x=0.3$ to 2.24 at $x=0.7$. In the twist-4 sector, the strange quark shows enormous scaling, with the $F_3^{K_{\bar{s}}}$ peak surging from 61.04 at $x=0.3$ to 258.52 at $x=0.7$, and the $H_3^{K_{\bar{s}}}$ minimum dropping drastically from $-4.83$ to $-39.46$. This strictly aligns with the kinematic expectation that the heavier strange quark naturally carries a larger fraction of the meson's longitudinal momentum, making highly localized, high-$x$ spatial configurations more probable for the $\bar{s}$ quark than for the $u$ quark. Notably, unlike the bimodal structure seen in the pion's unpolarized twist-2 distribution, both $F_1^{K_u}$ and $F_1^{K_{\bar{s}}}$ maintain strictly unimodal, centrally-peaked profiles across the evaluated domain, and the bimodal structure of pion is strictly suppressed in the kaon due to $\mathrm{SU}(3)$ flavor symmetry breaking through LFQM.

\subsection{Numerical results for longitudinal spatial distributions}

\begin{figure}[!ht]
    \centering
    \begin{minipage}{\linewidth}
        \includegraphics[width=\linewidth]{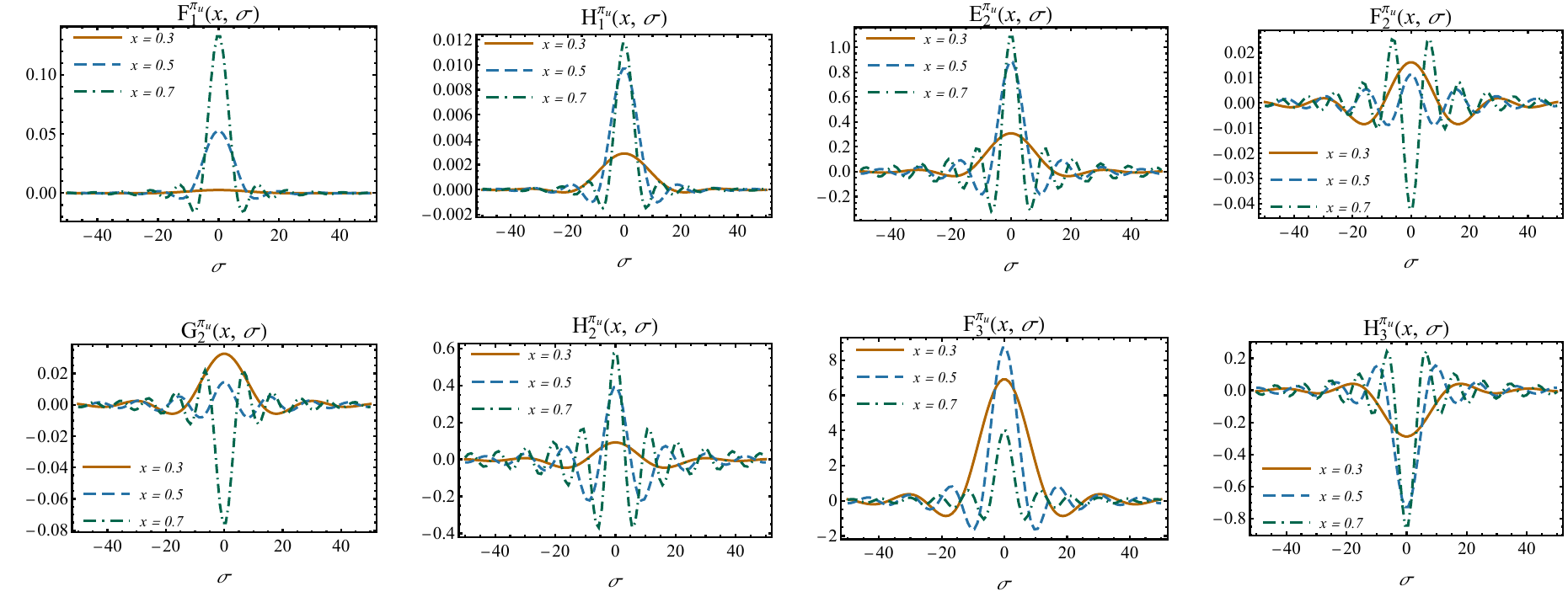}
    \end{minipage}

    \vspace{0.5cm}

    \begin{minipage}{\linewidth}
        \includegraphics[width=\linewidth]{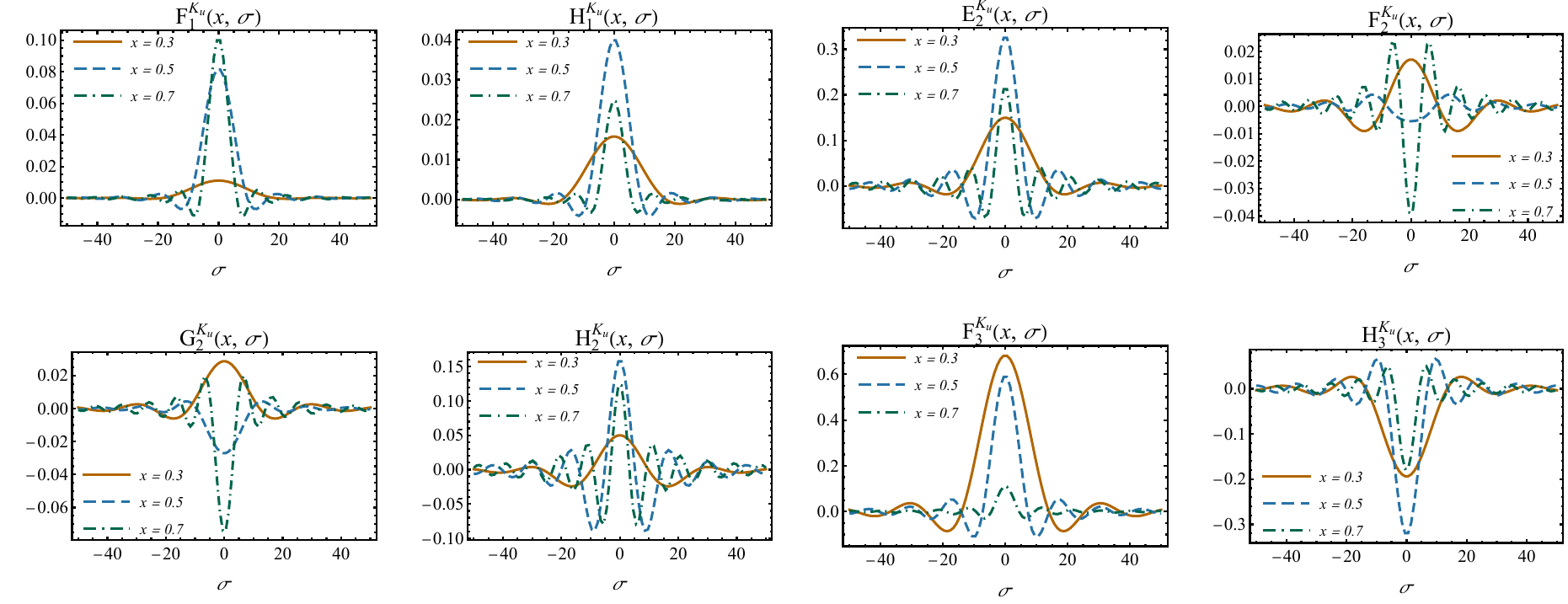}
    \end{minipage}

    \vspace{0.5cm}

    \begin{minipage}{\linewidth}
        \includegraphics[width=\linewidth]{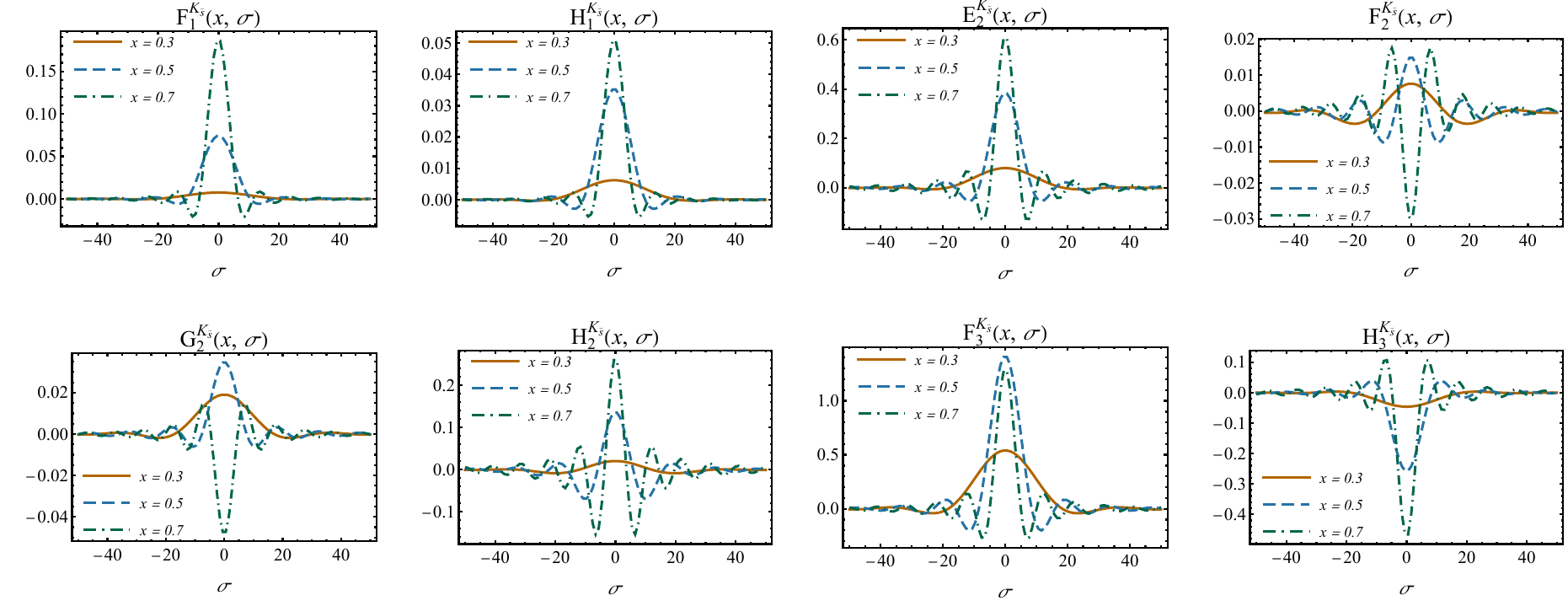}
    \end{minipage}

    \caption{Longitudinal spatial distributions ($\sigma$-space) for pion and kaon as a function of $\sigma$ at a fixed momentum transfer $t = -1.0$ GeV$^2$, shown at different $x$ values 0.3, 0.5 and 0.7 The first eight plots are for pion $u$ quark, the second eight plots are for kaon $u$ quark, and the last eight plots are for kaon $\bar{s}$ quark.}
    \label{fig:sigma_plots}
\end{figure}

In Fig.~\ref{fig:sigma_plots}, the longitudinal spatial distributions are plotted as a function of the conjugate coordinate $\sigma$ at $-t = 1.0$ GeV$^2$ for different values of $x$, particularly at $x = 0.3, 0.5$ and $0.7$. These are obtained by Fourier transforming the GPDs with respect to the skewness parameter $\xi$ at a fixed momentum transfer of $-t = 1.0$ GeV$^2$. Across all evaluated mesons and flavors, the most prominent characteristic is a distinct, diffraction-like oscillatory pattern. Mathematically, this pattern is a direct consequence of the finite integration bounds. Because the DGLAP kinematic domain restricts $\xi \le x$, the Fourier transform is truncated, naturally producing a sinc-like spatial distribution.

As the momentum fraction $x$ increases from 0.3 to 0.7, the allowable integration window for $\xi$ expands. Consequently, the central peak at $\sigma = 0$ systematically narrows, and the frequency of the peripheral spatial oscillations increases. For the pion $u$ quark, the central peaks of $F_1^{\pi_u}$ and $H_1^{\pi_u}$ grow from 0.003 to 0.13 and 0.01 respectively as $x$ increases to 0.7. The twist-3 $E_2^{\pi_u}$ and $H_2^{\pi_u}$ distributions also scale up to peaks of 1.08 and 0.57, respectively, while the $F_2^{\pi_u}$ and $G_2^{\pi_u}$ develop central minima at higher $x$. While the structural frequencies evolve similarly across quarks, the absolute amplitudes of these longitudinal distributions heavily depend on the specific flavor and twist, with twist-4 distributions again exhibiting the most significant magnitude scaling, highlighted by the pion's $F_3^{\pi_u}$ peak of 8.70 at $x=0.5$ and the $H_3^{\pi_u}$ minimum of $-0.84$ at $x=0.7$.

Comparing the longitudinal spatial profiles of the $u$ quark in the pion versus the kaon reveals the impact of the hadron mass scale, particularly within the higher-twist dynamics. While the twist-2 distributions $F_1$ and $H_1$ exhibit relatively similar central amplitudes between the two mesons, the twist-3 and twist-4 distributions undergo substantial kinematic suppression in the kaon compared to the nearly massless pion. For instance, at $x=0.5$, the twist-4 $F_3^{\pi_u}(x, \sigma)$ central peak for the pion increases to an amplitude of 8.70. In contrast, the identical distribution for the $u$ quark in the heavier kaon is suppressed, reaching a maximum amplitude of only 0.59. A similar disparity is visible in the twist-3 $E_2$ distribution, where the pion peak of 0.88 clearly dominates the corresponding kaon $u$ peak of 0.33. This order-of-magnitude suppression is a direct consequence of the heavier kaon mass $M_K$ entering the denominator of the covariant decomposition prefactors ($(P^+)^n / M^n$), mathematically enforcing the suppression of higher-twist correlations in heavier bound states.

Within the kaon itself, explicit $\mathrm{SU}(3)$ flavor symmetry breaking influences the longitudinal localization of the partons. Because the $\bar{s}$ quark is substantially heavier than the $u$ quark, it naturally resists spreading out in the longitudinal coordinate space, resulting in a sharper and denser localization near the central origin ($\sigma \approx 0$). At $x=0.7$, the twist-2 unpolarized distribution $F_1^{K_{\bar{s}}}$ reaches a central amplitude of 0.19, which is nearly double the peak density of the lighter $u$ quark ($F_1^{K_u} = 0.10$) within the same meson. This distinct mass hierarchy persists consistently throughout the higher-twist regime, with the $E_2$ and $F_3$ peaks of the strange quark scaling significantly higher than those of the $u$ quark. The remaining distributions, including $H_1$, $H_2$, and $H_3$, follow analogous scaling patterns. Specifically, the positive central peaks of $H_1$ and $H_2$ narrow with $x$, reaching 0.05 and 0.26 respectively for $H_1^{K_{\bar{s}}}$ and $H_2^{K_{\bar{s}}}$ at $x=0.7$, compared to 0.02 and 0.12 for $H_1^{K_u}$ and $H_2^{K_u}$. Furthermore, the central negative minimum of $H_3$ deepens considerably at higher momentum fractions, falling to $-0.47$ for $H_3^{K_{\bar{s}}}$ and $-0.18$ for $H_3^{K_u}$. Interestingly, despite these amplitude differences driven by constituent mass, the topological transitions remain consistent. The negative central inversions that develop in the $F_2$ and $G_2$ distributions at $x=0.7$ occur simultaneously across the pion $u$, kaon $u$ (minima of $-0.04$ and $-0.07$, respectively), and kaon $\bar{s}$ (minima of $-0.03$ and $-0.05$, respectively) profiles, indicating a shared kinematic behavior at high $x$.

\section{Summary}

In this work, we have presented a comprehensive numerical investigation into the multidimensional internal structure of the pion and kaon by evaluating their complete set of eight generalized parton distributions (GPDs). By employing the light-front quark model (LFQM) alongside the Brodsky-Huang-Lepage (BHL) prescription for the light-front wave functions (LFWFs), we extended the structural mapping beyond the commonly studied forward limit into the non-zero skewness ($\xi \neq 0$) domain. Our analysis covered leading-twist (twist-2) to higher-twist (twist-4) distributions.

Through our momentum-space analysis, we observed that all GPDs monotonically decay with increasing invariant momentum transfer $-t$, consistent with the finite spatial extent of hadronic form factors. A comparative study between the pion and kaon explicitly highlighted $\mathrm{SU}(3)$ flavor symmetry breaking. The heavier $\bar{s}$ quark in the kaon carries a larger fraction of the meson's longitudinal momentum, shifting its probability peaks to higher $x$ values relative to the lighter $u$ quarks. However, recent lattice and continuum QCD studies \cite{Chang:2025lrc, LatticeParton:2022zqc, RQCD:2019osh, Xu:2024nzp, Serna:2018dwk, Sultan:2018tet} indicate a much milder spatial skewing, suggesting that the extreme asymmetry seen here is a model-dependent consequence of this specific mass prescription rather than a universal QCD feature. Crucially, our evaluations at varying skewness numerically validated the hermiticity constraints imposed by the underlying generalized transverse momentum dependent distributions (GTMDs). We confirmed that the twist-3 chiral distributions $F_2$ and $H_2$ possess strict odd symmetry and strictly vanish at $\xi=0$, precluding them from having direct TMD analogs.

To construct a 3D tomographic picture of the partons, we performed Fourier transforms of the GPDs into both transverse and longitudinal coordinate spaces. In the transverse plane, the impact parameter dependent parton distribution functions (IPDPDFs) revealed unique topological structures, such as the emergence of a bimodal, ring-like configuration for highly energetic unpolarized quarks ($F_1^{\pi_u}$) and substantial spatial amplitude amplification for higher twists ($F_3^{\pi_u}$ and $H_3^{\pi_u}$). In the longitudinal $\sigma$-space, the truncated integration over the physical DGLAP kinematic domain naturally generated diffraction-like oscillatory patterns.

Ultimately, contrasting the spatial tomography of the pion against the kaon demonstrated that higher-twist spatial correlations are severely dampened by the larger macroscopic mass of heavier bound states. These findings provide critical theoretical inputs for understanding non-perturbative QCD dynamics and offer valuable baseline initial conditions for DGLAP evolution relevant to future exclusive scattering experiments.

\begin{acknowledgements}
    H.D. would like to thank the Science and Engineering Research Board, AnusandhanNational Research Foundation, Government of India under the scheme SERB-POWER Fellowship (Ref No. SPF/2023/000116) for financial support.
\end{acknowledgements}

\bibliographystyle{apsrev4-2}
\bibliography{references}

\end{document}